\documentclass[a4paper,11pt]{article}
\pdfoutput=1 
\usepackage{jheppub} 
\usepackage[T1]{fontenc} 
\usepackage{amsmath,amssymb,epsfig,txfonts,relsize}
\usepackage[utf8]{inputenc}

\allowdisplaybreaks

\definecolor{nicered}{rgb}{0.7,0.1,0.1}
\definecolor{nicegreen}{rgb}{0.1,0.5,0.1}
\definecolor{niceblue}{rgb}{0.0,0.1,0.7}
\hypersetup{colorlinks,citecolor=nicegreen,linkcolor=nicered,urlcolor=niceblue}

\def\bm#1{\mbox{\boldmath$#1$\unboldmath}}      
\def \beq{\begin{equation}}
\def \eeq{\end{equation}}
\def \bea{\begin{eqnarray}}
\def \eea{\end{eqnarray}}

\title{Constraints on the quartic Higgs self-coupling from double-Higgs production at future hadron colliders}

\author[1,2,5]{Wojciech Bizo\'{n},}

\author[3,5,6]{Ulrich Haisch,}

\author[4,5]{and Luca Rottoli}

\affiliation[1]{Institut f{\"u}r Theoretische Teilchenphysik (TTP), KIT, 76128 Karlsruhe, Germany}  

\affiliation[2]{Institut f{\"u}r Kernphysik (IKP), KIT, 76344 Eggenstein-Leopoldshafen, Germany}  

\affiliation[3]{Max Planck Institute for Physics, F{\"o}hringer Ring 6,  80805 M{\"u}nchen, Germany}   

\affiliation[4]{Dipartimento di Fisica G. Occhialini, U2, Universit{\`a} degli Studi di Milano-Bicocca, \\ Piazza della Scienza, 3, 20126 Milano, Italy}  

\affiliation[5]{Rudolf Peierls Centre for Theoretical Physics, University of Oxford, \\ OX1 3NP Oxford, United Kingdom}
   
\affiliation[6]{CERN, Theoretical Physics Department,  CH-1211 Geneva 23, Switzerland}

\emailAdd{wojciech.bizon@kit.edu}
\emailAdd{haisch@mpp.mpg.de}
\emailAdd{luca.rottoli@unimib.it}

\abstract{
\phantom{iii} We study the indirect constraints on the quartic Higgs self-coupling that arise from double-Higgs production at future hadron colliders. To this purpose, we calculate the two-loop contributions to the $gg \to hh$ amplitudes that involve a modified $h^4$ vertex. Based on our  results, we estimate~the~reach of a $pp$ collider operating at $27 \, {\rm TeV}$ and $100 \, {\rm TeV}$ centre-of-mass energy in constraining the cubic and quartic Higgs self-couplings by  measurements of double-Higgs and triple-Higgs production in gluon-fusion. 
}

\preprint{CERN-TH-2018-218, OUTP-17-21P}

\begin{document} 

\maketitle

\section{Introduction}
\label{sec:introduction}

In the standard model (SM) of elementary particle physics, the interactions involving the Higgs boson take the following simple form 
\begin{equation} \label{eq:LHiggs}
{\cal L} \supset \left |D_\mu H \right |^2 - \sum_f \left ( y_f \hspace{0.25mm} \bar f_L H f_R + {\rm h.c.} \right )  - V \,, \qquad V = -\mu \left |H \right |^2 + \lambda \left |H \right |^4 \,,
\end{equation}
where $D_\mu$ denotes the $S\hspace{-0.5mm}U(2)_L \times U(1)_Y$ covariant derivative, $H$ is the Higgs doublet, the subscripts~$L,R$ indicate the chirality of fermionic fields and $y_f$ are the so-called Yukawa couplings.

An  obvious question that one may ask concerning (\ref{eq:LHiggs}) is: what is presently known about the interactions of the Higgs boson? The ATLAS and CMS combination of the LHC~Run-I Higgs measurements~\cite{Khachatryan:2016vau} imply that the gauge-Higgs interactions, which are encoded by~$|D_\mu H|^2$, agree with the SM predictions at the level of $10\%$. The Yukawa interactions $y_f \hspace{0.25mm} \bar f_L H f_R + {\rm h.c.}$, on the other hand, have been tested with this accuracy only in the case of the tau lepton, while the constraints on the top and bottom Yukawa couplings just reach the $20\%$ level. Apart from the muon Yukawa coupling, which is marginally constrained by the existing LHC data, first and second generation Yukawa couplings are not directly probed at present. In the case of the Higgs potential~$V$, the vacuum expectation value (VEV) of $H$ is known since the discovery of the $W$ and $Z$ bosons, while the LHC discovery of a scalar with a  mass of around $125 \, {\rm GeV}$  tells us about the second derivative of $V$ around its VEV, because this quantity determines the Higgs mass. The $h^3$ and $h^4$ Higgs self-interactions that result from (\ref{eq:LHiggs}) are in contrast essentially untested at the moment.

\begin{figure}[!t]
\begin{center}
\includegraphics[width=0.975 \textwidth]{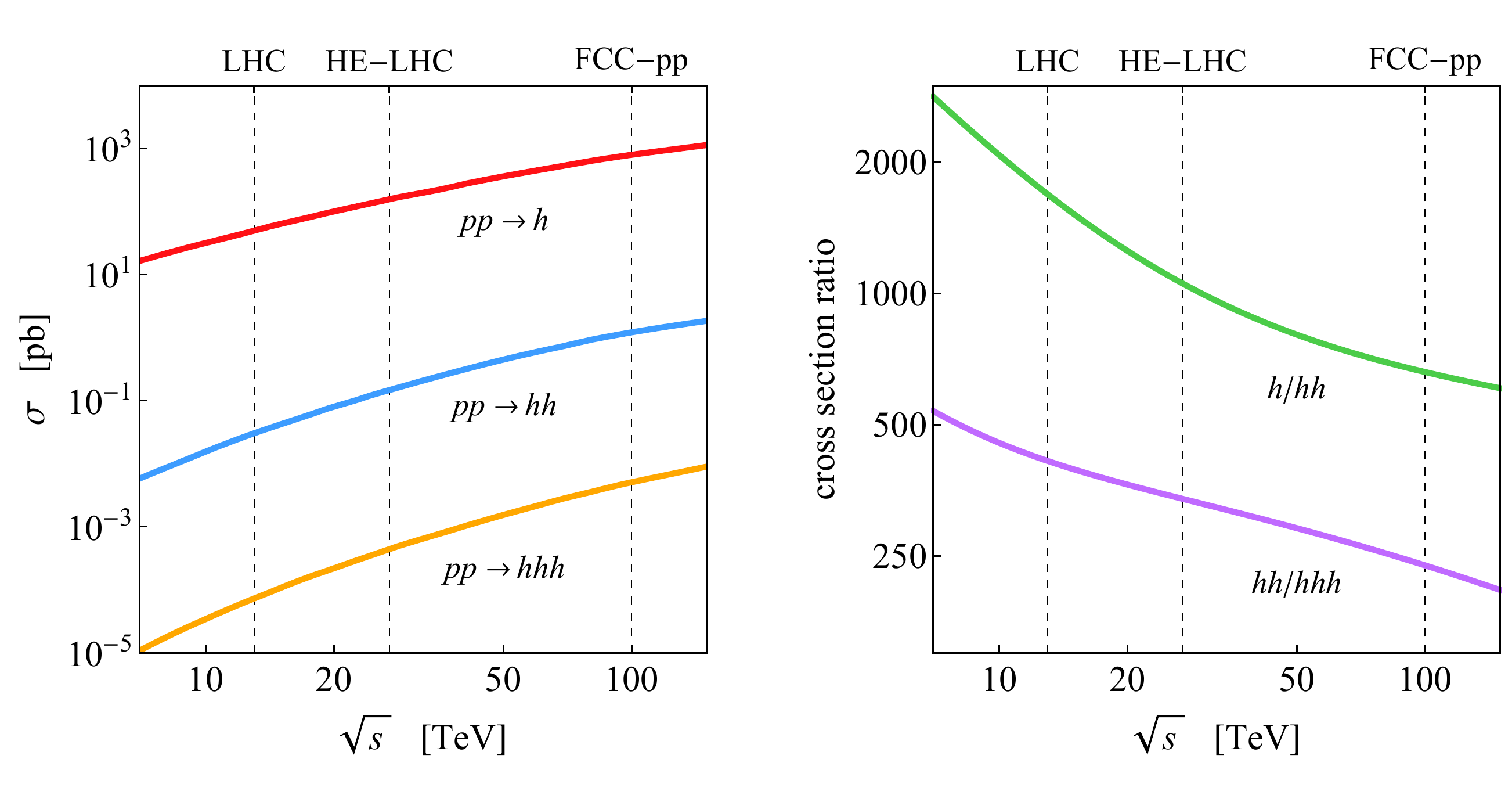} 
\vspace{0mm}
\caption{\label{fig:multihsxsec} Left: Total production  cross section for $pp \to h$ (red),  $pp \to hh$ (blue) and  $pp \to hhh$~(yellow) as a function of $\sqrt{s}$. Right:  Dependence of the cross section ratio $\sigma (pp \to h)/\sigma (pp \to hh)$ (green) and $\sigma (pp \to hh)/\sigma (pp \to hhh)$ (purple) on the  collider CM energy. The shown predictions are based on the state-of-the-art  SM calculations of single-Higgs~\cite{Anastasiou:2015ema,Anastasiou:2016cez,Contino:2016spe}, double-Higgs~\cite{Borowka:2016ehy, Borowka:2016ypz,Heinrich:2017kxx,Grazzini:2018bsd} and triple-Higgs~\cite{Maltoni:2014eza} production.} 
\end{center}
\end{figure}

Given our limited knowledge of  the properties of the discovered $125 \, {\rm GeV}$ resonance, constraining or measuring as many of its so far poorly known or unknown couplings will be an important part of any future high-energy programme. In the case of the Higgs self-couplings  the most obvious way to get access to the cubic and quartic  interactions consists in searching for multi-Higgs production. Unfortunately, all multi-Higgs production rates are quite small in the SM, as can be seen from~Figure~\ref{fig:multihsxsec},  making already  LHC measurements of double-Higgs production  a formidable task.  As a result, at best ${\cal O} (1)$ determinations of the cubic Higgs self-coupling  seem to be possible at the LHC~(cf.~for instance~\cite{Goertz:2014qta,Azatov:2015oxa,ATL-PHYS-PUB-2015-046,Kling:2016lay,ATL-PHYS-PUB-2016-024,DiVita:2017eyz,ATL-PHYS-PUB-2017-001}). Significantly improved prospects in extracting the $h^3$ coupling would be offered by a high-energy upgrade of the LHC~(HE-LHC) to $27 \, {\rm TeV}$~\cite{Goncalves:2018qas} or a future circular collider~(FCC-pp) operating at a centre-of-mass~(CM)~energy of $100 \, {\rm TeV}$~\cite{Azatov:2015oxa,Barr:2014sga,He:2015spf,Contino:2016spe,Mangano:2016jyj,Banerjee:2018yxy,Chang:2018uwu}. A~$100 \, {\rm TeV}$ $pp$ machine, in particular, may ultimately allow one to determine the cubic Higgs self-coupling with a statistical precision of the order of a few percent. Even a $100 \, {\rm TeV}$ FCC-pp collider is, however, not powerful enough to determine  the SM triple-Higgs production rate to an accuracy better than just order one~\cite{Contino:2016spe,Mangano:2016jyj,Papaefstathiou:2015paa,Chen:2015gva,Fuks:2015hna,Kilian:2017nio,Fuks:2017zkg}. The resulting bounds on the quartic Higgs self-coupling turn out to be weak, in general allowing for ${\cal O} (10)$ modifications of the $h^4$ vertex with respect to the~SM prediction. 

Motivated by the above observations,  we apply  in this work  the  general idea of testing the~$h^3$ interaction indirectly~\cite{McCullough:2013rea,Gorbahn:2016uoy,Degrassi:2016wml,Bizon:2016wgr, Degrassi:2017ucl,Kribs:2017znd,DiVita:2017eyz,Maltoni:2017ims,DiVita:2017vrr, Maltoni:2018ttu,Liu:2018peg,Borowka:2018pxx,Gorbahn:2019lwq} to the case of the $h^4$ vertex. Specifically, we consider the constraints on the quartic Higgs self-coupling that future precision measurements of double-Higgs production in gluon-fusion may provide. In order to determine the dependence of the $gg \to hh$ distributions on the value of the $h^4$ coupling, we calculate the relevant electroweak~(EW) two-loop amplitudes  and combine them with the exact ${\cal O} (\alpha_s^2)$ matrix elements~\cite{Borowka:2016ehy,Borowka:2016ypz,Heinrich:2017kxx}. This allows us to predict the cross section and various distributions for double-Higgs production at the next-to-leading~order~(NLO) in QCD, including arbitrary modifications of the cubic and quartic Higgs self-couplings.  We then perform an exploratory study of the synergy and complementarity  of double-Higgs and triple-Higgs production in constraining the $h^3$ and $h^4$ interactions, considering both the HE-LHC and a FCC-pp machine as an example. A similar analysis has  very recently also been performed~in~\cite{Borowka:2018pxx}.  For completeness, we add that the indirect constraints on the quartic Higgs self-coupling that high-energy $e^+ e^-$ machines may be able to set have been studied in~\cite{Maltoni:2018ttu,Liu:2018peg}.  In these articles it has been shown that future lepton colliders can in general only provide  coarse bounds on possible modifications of the $h^4$ vertex, if one makes no assumption about how ultraviolet~(UV) physics alters the cubic and quartic Higgs self-interactions. We will compare our limits to those obtained in the publications~\cite{Maltoni:2018ttu,Liu:2018peg,Borowka:2018pxx}. 

This  article is structured as follows. In Section~\ref{sec:preliminaries} we introduce our parameterisation of the Higgs potential and discuss how the coefficients entering it are related 
to the Wilson coefficients of two higher-dimensional operators of the SM effective field theory (SMEFT). The calculation of the two-loop corrections to the $g g \to hh$ scattering amplitude resulting from a modified quartic Higgs self-coupling is described in Section~\ref{sec:calculation}. In Section~\ref{sec:numerics} we determine the hypothetical reach of a $27 \, {\rm TeV}$ HE-LHC and a $100 \, {\rm TeV}$ FCC-pp in constraining the cubic and quartic Higgs self-couplings by  measurements of double-Higgs and triple-Higgs production in gluon-fusion.  Section~\ref{sec:conclusions} contains our conclusions. 

\section{Preliminaries}
\label{sec:preliminaries}

After EW symmetry breaking, the cubic and quartic self-interactions of the Higgs field $h$ can be parameterised  in a model-independent fashion by 
\beq \label{eq:V}
V \supset \kappa_3 \hspace{0.25mm} \lambda v h^3 + \kappa_4 \hspace{0.25mm} \frac{\lambda}{4} \hspace{0.25mm}  h^4 \,.
\eeq
Here $\lambda  = m_h^2/(2v^2)$ with $m_h \simeq 125 \, {\rm GeV}$ the Higgs-boson mass and $v \simeq 246 \, {\rm GeV}$ the EW VEV. Notice that the normalisation of the terms  in~(\ref{eq:V}) has been chosen such that within the SM one has~$\kappa_{3} = \kappa_{4} = 1$. 

In the presence of physics beyond the SM~(BSM) the coefficients $\kappa_3$ and $\kappa_4$ will in general deviate from 1. As an illustrative example, let us consider the following two  terms 
\beq \label{eq:LSMEFT}
{\cal L}_{\rm SMEFT}\supset {\cal O}_6 + {\cal O}_8 = -\frac{\lambda \hspace{0.5mm} \bar c_6}{v^2} \hspace{0.5mm}  \left |H \right |^6 -\frac{\lambda \hspace{0.5mm} \bar c_8}{v^4} \hspace{0.5mm}  \left |H \right |^8 \,,
\eeq
in the SMEFT. In such a case  the  parameters $\kappa_3$ and~$\kappa_4$ are related at tree level to the Wilson coefficients $\bar c_6$ and $\bar c_8$ via 
\beq \label{eq:kappa3kappa4}
\kappa_3= 1 + \Delta \kappa_{3} = 1 + \bar c_6 + 2 \hspace{0.25mm} \bar c_8  \,,  \qquad 
\kappa_4 = 1 + \Delta \kappa_{4} = 1 + 6 \hspace{0.25mm} \bar c_6 + 16 \hspace{0.25mm}  \bar c_8  \,.
\eeq
The relations~(\ref{eq:kappa3kappa4}) imply that if the dimension-six operator ${\cal O}_6$ represents the only numerically relevant modification in the SMEFT, the shifts in the cubic and quartic Higgs self-couplings are strongly correlated as they obey $\Delta \kappa_4  = 6 \Delta \kappa_3$. This correlation is, however, broken by the dimension-eight contribution ${\cal O}_8$, if this operator receives a non-zero Wilson coefficient. Notice that the initial conditions of the Wilson coefficients $\bar c_6$ and $\bar c_8$ can be found  in any UV complete BSM model by a suitable matching calculation. If the new interactions that lead to ${\cal O}_6$ and ${\cal O}_8$ are weakly coupled and the new-physics scale $\Lambda$ is in the TeV range, one expects on general grounds that the dimension-eight and dimension-six contributions have the following hierarchy~$\bar c_8/\bar c_6 = {\cal O} (v^2/\Lambda^2) \ll 1$. The Wilson coefficients $\bar c_8$ and $\bar c_6$ can however be  of the same order of magnitude if the underlying UV theory is strongly coupled or the new-physics scale $\Lambda$ is at (or not far~above) the~EW scale. To~achieve the inverted hierarchy $\bar c_8/\bar c_6 \gg 1$ the new particles that give rise to (\ref{eq:LSMEFT})  have to have masses of ${\cal O} (v)$ and have to have interactions with the Higgs doublet $H$ that are strong --- SM extensions with  colourless $S\!U(2)$ quadruplets $\Theta$~\cite{deBlas:2014mba} can for instance lead to such an  inverted hierarchy if  the quadruplet  is sufficiently light and the Higgs portal coupling $|\Theta|^2 \, |H|^2$ is sufficiently large.  

In our work we choose to be agnostic about how UV dynamics modifies the Higgs self-interactions, and hence make no assumption about the actual size of $\bar c_6$ and $\bar c_8$. In this case, the cubic and quartic Higgs self-couplings can  deviate independently from the SM predictions.  The important point is now that even if $\Delta \kappa_{3}$ and $\Delta \kappa_{4}$  are treated as free parameters, quantum processes such as $gg \to h$ or loop corrections to $e^+ e^-  \to hhZ$ can still be calculated consistently  as long as the SMEFT  is used to perform the computations~(see~\cite{McCullough:2013rea,Gorbahn:2016uoy,Degrassi:2016wml,Bizon:2016wgr, Degrassi:2017ucl,Kribs:2017znd,Maltoni:2018ttu,Liu:2018peg,Borowka:2018pxx,Gorbahn:2019lwq}  for non-trivial one-loop and two-loop examples and further explanations). Since  modifications in the cubic and quartic Higgs self-coupling are most commonly parametrised by $\Delta \kappa_{3}$ and $\Delta \kappa_{4}$, we will also use this parameterisation in what follows. We however emphasise that constraints on the latter parameters can always be translated into bounds on the Wilson coefficients $\bar c_6$ and $\bar c_8$ by means of~(\ref{eq:kappa3kappa4}).  

\section{Calculation}
\label{sec:calculation}

The scattering amplitude describing the process $g(p_1) + g(p_2) \to h(p_3) + h(p_4)$ can be written as 
\beq \label{eq:Agghh}
{\cal A} \left (gg\to hh \right ) = \delta^{a_1 a_2}  \hspace{0.5mm} \epsilon_{1}^{\mu} (p_1) \hspace{0.5mm} \epsilon_{2}^\nu (p_2) \hspace{1mm} {\cal A}_{\mu \nu} \,,
\eeq
where $a_{1}$ and $a_{2}$ denote colour indices while $\epsilon_{1}^\mu (p_{1})$ and $\epsilon_{2}^\nu (p_{2})$ are the polarisation vectors of the two initial-state gluons. Using Lorentz symmetry, parity conservation and gauge invariance, one can show that the amplitude tensor ${\cal A}_{\mu \nu}$ that appears in~(\ref{eq:Agghh}) can be expressed in terms of two form factors as follows 
\beq \label{eq:Amunu}
{\cal A}_{\mu \nu} =  \sum_{m=1}^2 T_{m\, \mu \nu} \, {\cal F}_m   \,,
\eeq
where~\cite{Glover:1987nx}
\beq  \label{eq:Tmmunu}
\begin{split}
T_{1\, \mu \nu} & = \eta_{\mu \nu} - \frac{p_{1 \, \nu} \hspace{0.5mm} p_{2 \, \mu}}{p_1 \cdot p_2} \,, \\[2mm]
T_{2\, \mu \nu} & = \eta_{\mu \nu} + \frac{1}{p_T^2 \left ( p_1 \cdot p_2 \right)} \, 
\left ( \, m_h^2 \hspace{0.5mm}  p_{1 \, \nu} \hspace{0.5mm}  p_{2 \, \mu} 
- 2 \left ( p_1 \cdot p_3 \right ) p_{2 \, \mu}  \hspace{0.5mm}  p_{3 \, \nu}  \right. \\[1mm]
& \hspace{3.4cm} \left . - 2 \left ( p_2 \cdot p_3 \right )  p_{1 \, \nu}  \hspace{0.5mm}  p_{3 \, \mu} 
+ 2 \left ( p_1 \cdot p_2 \right )  p_{3 \, \mu} \hspace{0.5mm}   p_{3 \, \nu} \, \right ) \,,
\end{split}
\eeq
with $\eta_{\mu \nu} = {\rm diag} \left ( 1, -1, -1, -1 \right)$ the Minkowski metric and $p_T$ denoting the Higgs transverse momentum. In terms of the partonic Mandelstam variables 
\beq \label{eq:mandelstam}
\hat s = \left (p_1 + p_2 \right )^2 \,, \qquad 
\hat t = \left (p_1 - p_3 \right )^2 \,, \qquad 
\hat u = \left (p_2 - p_3 \right )^2 \,,
\eeq
one can write 
\beq \label{eq:pT2}
p_T^2 = \frac{\hat t \hspace{0.75mm} \hat u - m_h^4}{\hat s} \,.
\eeq
Notice furthermore that $p_1^2=p_2^2 = 0$ while $p_3^2=p_4^2 = m_h^2$ and that the Mandelstam variables fulfill the relation $\hat s + \hat t + \hat u = 2 m_h^2$. 

The form factors entering~(\ref{eq:Amunu}) are most conveniently extracted by using a projection procedure. The appropriate projectors read~(see~\cite{Borowka:2016ehy} for example) 
\beq \label{eq:projectors}
\begin{split}
P_1^{\mu \nu} & = \frac{1}{4 \left (d-3 \right)} \, \left [  \, \left ( d-2 \right )   T_1^{\mu \nu} -  \left ( d-4 \right )  T_2^{\mu \nu} \, \right ]  \,, \\[2mm]
P_2^{\mu \nu} & =  \frac{1}{4 \left (d-3 \right)} \, \left [  \, -\left ( d-4 \right )   T_1^{\mu \nu} +  \left ( d-2 \right )  T_2^{\mu \nu} \, \right ]  \,, 
\end{split}
\eeq
where $d = 4 - 2 \epsilon$ denotes the number of space-time dimensions. 

\begin{figure}[!t]
\begin{center}
\includegraphics[width=0.8 \textwidth]{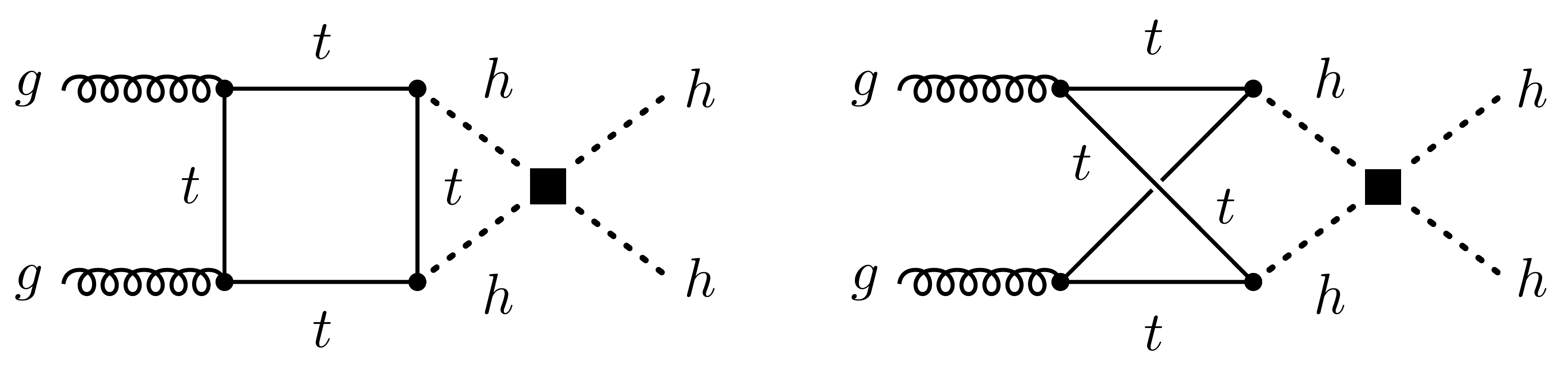} 
\vspace{4mm}
\caption{\label{fig:diagrams} Prototypes of two-loop Feynman diagrams with an insertion of an effective quartic Higgs self-coupling~(black square) that contribute to the process $gg \to hh$.} 
\end{center}
\end{figure}

After applying the projectors~(\ref{eq:projectors}) each of the two $gg \to hh$ form factors can be calculated separately. Since the form factors are independent of the external polarisation vectors, all the standard techniques employed in multi-loop computations can be applied.  In practice, we proceed in the following way. We generate the relevant  two-loop Feynman diagrams with {\tt FeynArts}~\cite{Hahn:2000kx}. Representative examples of two-loop graphs are shown in Figure~\ref{fig:diagrams}. The projection onto form factors as well as  the colour  and Dirac algebra is performed with the help of~{\tt FORM}~\cite{Vermaseren:2000nd}. The resulting two-loop integrals are then evaluated numerically using the {\tt pySecDec}~\cite{Borowka:2012yc,Borowka:2015mxa,Borowka:2017idc} package. Including all two-loop diagrams leads to UV-finite results for the form factors, and we have checked that  the  double and single $1/\epsilon$ poles cancel to a relative accuracy of at least a permyriad for all calculated phase-space points.  Since in addition the quartic Higgs self-coupling does not result in  a finite one-loop correction of the Higgs wave function, it follows that the calculation of the ${\cal O} (\kappa_4)$ contributions to the $gg \to hh$ form factors  arising from the Feynman diagrams depicted in Figure~\ref{fig:diagrams} does not require renormalisation. 

As a further check  of our numerical results, we have performed a systematic expansion of the  two-loop form factors in the limit $m_t^2 \gg m_h^2, \hat s, \hat t, \hat u$ by employing  the method of asymptotic expansions~(see~\cite{Smirnov:2002pj}~for a review and~\cite{Gorbahn:2019lwq} an application in a similar context). Our analytic calculation made use of~{\tt MATAD}~\cite{Steinhauser:2000ry},~{\tt LiteRed}~\cite{Lee:2013mka}, the tensor reduction procedures described in~\cite{Tarasov:1995jf,Tarasov:1996br,Tarasov:1997kx} and the  results of  massive two-loop vacuum integrals first given in~\cite{Avdeev:1994db}. The agreement of the final results in the limit $\hat s < m_t^2$ between the two approaches serves as a non-trivial cross-check of our numerical computations.

\begin{figure}[!t]
\begin{center}
\includegraphics[width=0.875 \textwidth]{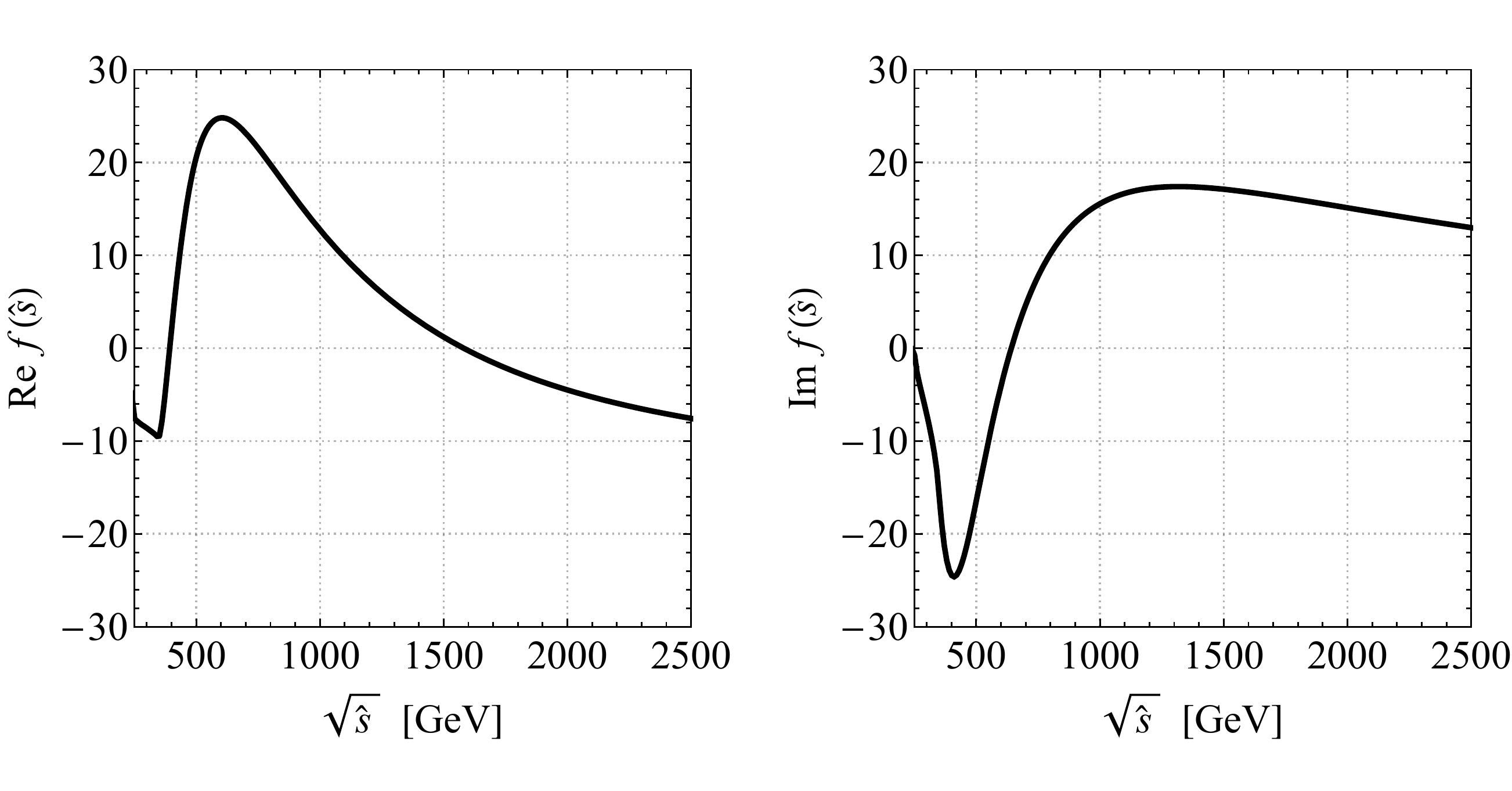} 

\vspace{-4mm}

\includegraphics[width=0.875 \textwidth]{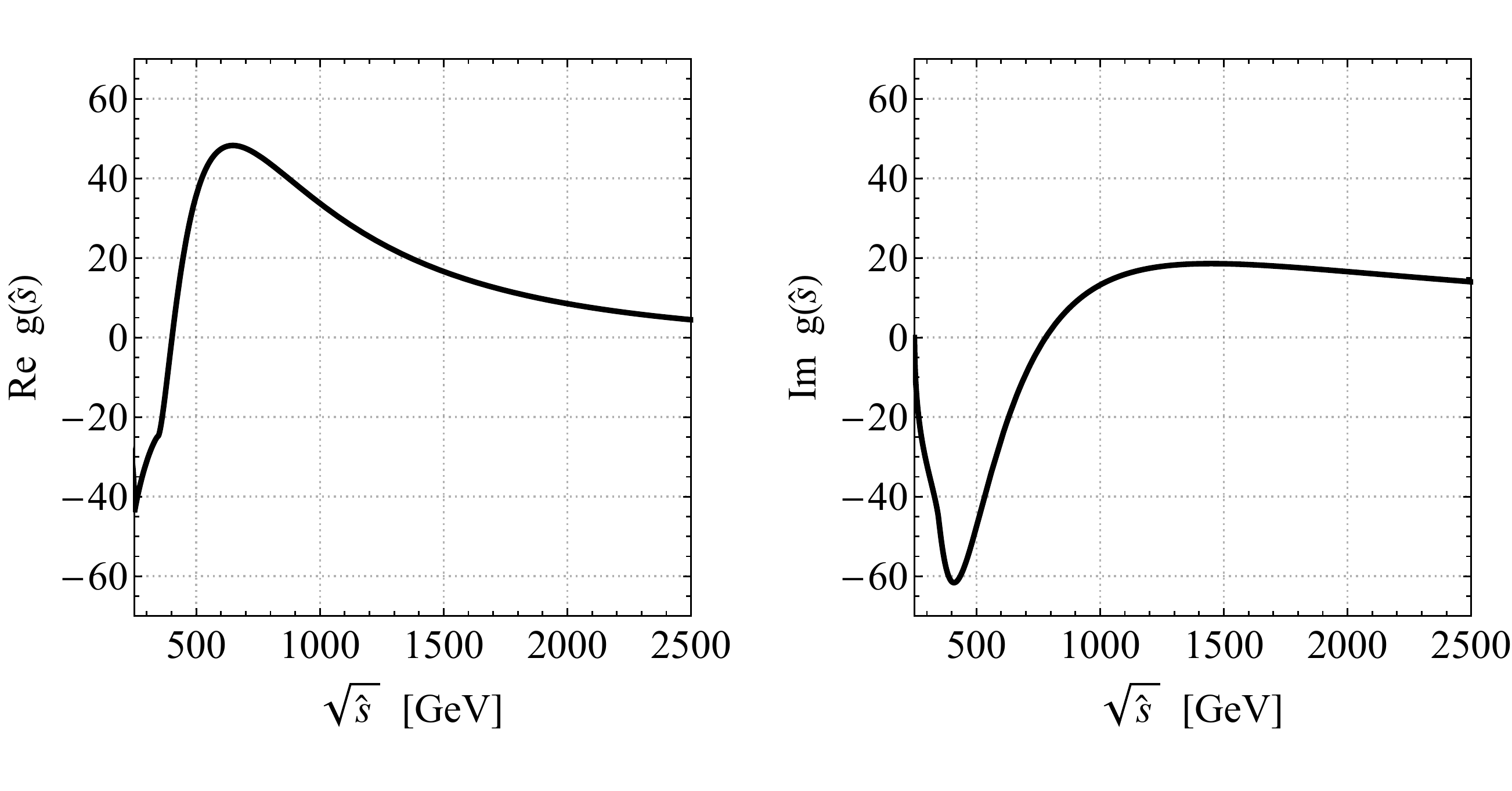} 
\vspace{-4mm}

\includegraphics[width=0.875 \textwidth]{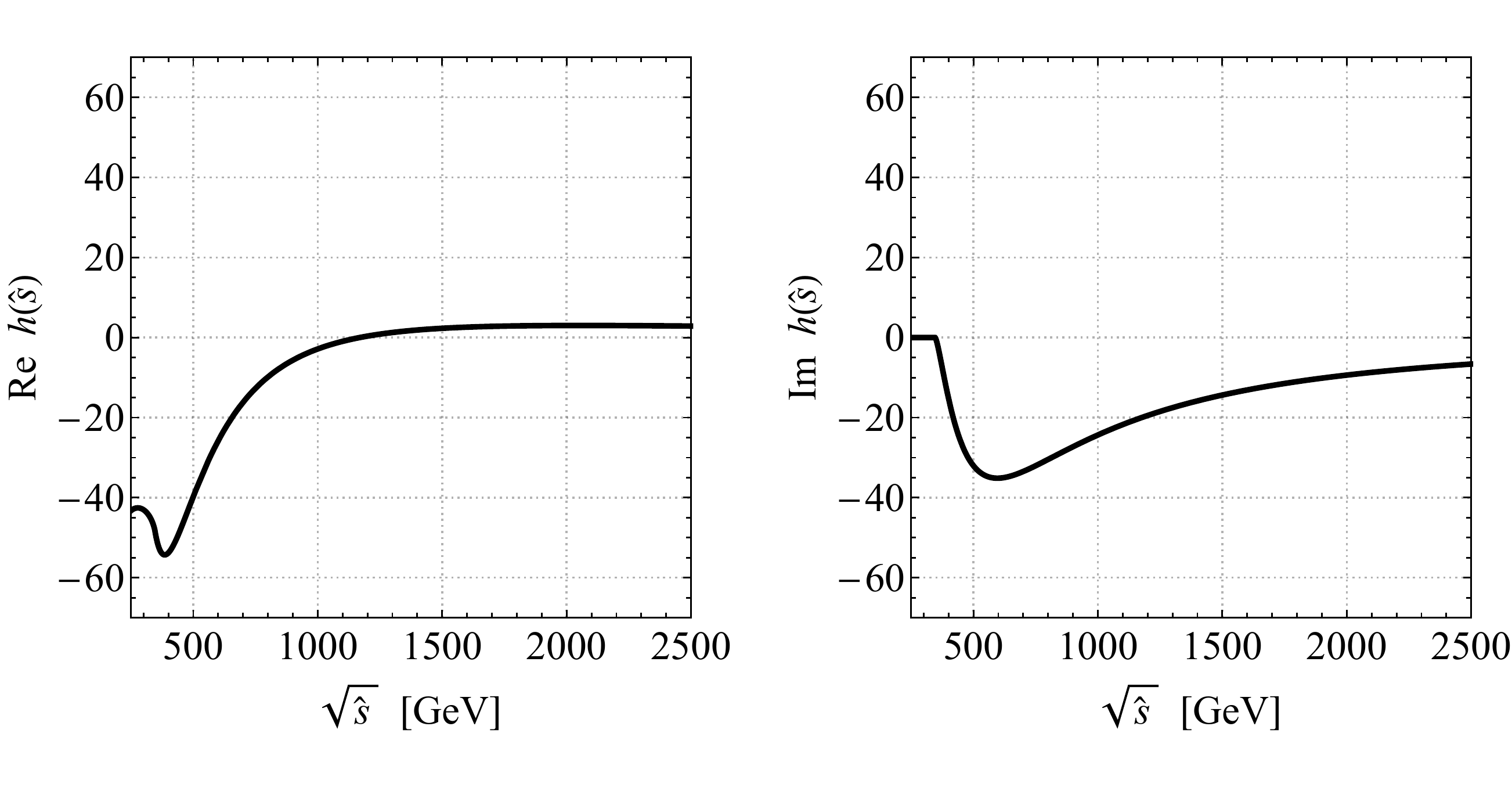} 
\vspace{-4mm}
\caption{\label{fig:refimf} Real part (left panel) and imaginary part (right panel) of the function $f (\hat s)$ (upper row), $g (\hat s)$~(middle row) and $h (\hat s)$~(lower row) as introduced in~(\ref{eq:DeltaF1F2}), (\ref{eq:gfunction}) and  (\ref{eq:hfunction}), respectively.} 
\end{center}
\end{figure}

Using the results of our numerical two-loop calculation, we find that the ${\cal O} (\kappa_4)$ corrections to the two $gg \to hh$ form factors that arise from the graphs shown in Figure~\ref{fig:diagrams} can be written as 
\beq \label{eq:DeltaF1F2}
\Delta {\cal F}_{1,n}^{\kappa_4} = \frac{\alpha_s}{4 \pi} \frac{\lambda  \, \kappa_4}{(4 \pi)^2} \; y_t^2 \, f ( \hat s  ) \,, \qquad 
\Delta {\cal F}_{2,n}^{\kappa_4} = 0 \,.
\eeq
Here $\alpha_s = g_s^2/(4 \pi)$ is the strong coupling constant, $y_t = \sqrt{2} m_t/v$ denotes the SM top-quark Yukawa coupling and the subscript $n$ indicates that the above corrections arise from non-factorisable two-loop diagrams. Two features of the expressions~(\ref{eq:DeltaF1F2}) are worth emphasising. First, the correction to the spin-0 form factor~${\cal F}_1$ depends only on $\hat s$ but not on the other two Mandelstam variables~$\hat t, \hat u$ $\big($or the combination $p_T^2$ introduced in~(\ref{eq:pT2})$\big)$. Second, the correction to the spin-2 form factor ${\cal F}_2$ turns out to be identical to zero.  The first feature is readily understood by noticing that the momentum routing in the two  diagrams in~Figure~\ref{fig:diagrams} can be chosen such that the external momenta only enter in the combination~$p_1 + p_2$. Due to Lorentz invariance the corresponding Feynman integrals can thus only depend on $\hat s = (p_1+p_2)^2 = 2 p_1 \cdot p_2$. The vanishing of the correction $\Delta {\cal F}_{2,n}$ to the spin-2 form factor follows for instance from the observation that the vertex $h^4$ can effectively be generated via the $s$-channel exchange of a heavy scalar $S$ that interacts with the Higgs boson through a term of the form $S h^2$. As a result the graphs in Figure~\ref{fig:diagrams} are mathematically equivalent to the off-shell production of a heavy CP-even spin-0 state that subsequently decays to $hh$. The corresponding scattering amplitude has evidently no spin-2 component. 

The  real and imaginary parts of the function $f(\hat s)$ that appears in~(\ref{eq:DeltaF1F2}) are  depicted in the upper row of Figure~\ref{fig:refimf}. The shown results correspond to $m_h = 125 \, {\rm GeV}$ and $m_t = 173 \, {\rm GeV}$. From the left panel one sees that  the real part of~$f(\hat s)$ changes its slope at the top-quark threshold $\sqrt{\hat s} = 2 m_t \simeq 375 \, {\rm GeV}$ and has a pronounced global maximum at $\sqrt{\hat s} = 2 \left ( m_h + m_ t \right ) \simeq 600 \, {\rm GeV}$. As illustrated on the right-hand side and expected from the optical theorem, the imaginary part of~$f(\hat s)$  instead vanishes at the  threshold for double-Higgs production,~i.e.~$\sqrt{\hat s} = 2 m_h \simeq 250 \, {\rm GeV}$.  It  then decreases rapidly, developing a distinct minimum in the vicinity of $\sqrt{\hat s} \simeq 400 \, {\rm GeV}$. We will see in Section~\ref{sec:numerics2} that the non-trivial $\hat s$ dependence of the real and imaginary parts of~$f(\hat s)$ leads to distortions in the~kinematic $gg \to hh$ distributions such as the invariant mass $m_{hh}$ of the di-Higgs~system. 

\begin{figure}[!t]
\begin{center}
\includegraphics[width=0.8 \textwidth]{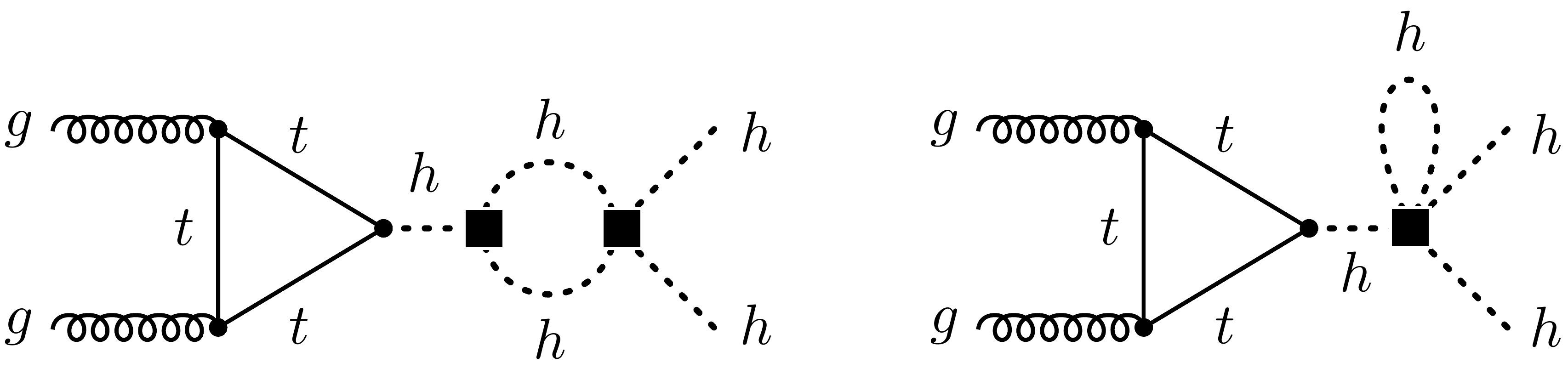} 
\vspace{4mm}
\caption{\label{fig:diagramsadditional} Two-loop graph with an effective cubic and quartic (left diagram) and quintic (right diagram) Higgs self-coupling that gives rise to $gg \to hh$ production. The effective interactions are indicated by black squares. For further details consult the text.} 
\end{center}
\end{figure}

Additional corrections to the $gg \to hh$ form factors~(\ref{eq:Amunu}) that are proportional to the self-coupling modifier $\kappa_4$ arise from the two types of Feynman diagrams displayed in Figure~\ref{fig:diagramsadditional}. Since the operators ${\cal O}_6$ and ${\cal O}_8$ do not generate couplings between two Higgs and two would-be Goldstone fields only graphs with  Higgs-boson  exchange contribute to the class of diagrams with a cubic and~a~quartic self-interaction. As a result this contribution is of ${\cal O} (\kappa_3 \kappa_4)$. In the case of the graphs with a quintic scalar self-coupling instead both Higgs and would-be Goldstone loops are present, and this contribution turns out to be proportional to ${\cal O} (\kappa_4)$. Besides the two-loop diagrams shown in the latter figure one-loop counterterm contributions associated to the renormalisation of the Higgs tadpole $T_h$, its wave function $Z_h$ and its mass $m_h$ as well as corrections associated to operator renormalisation have to be included to obtain a gauge-invariant result. We perform the renormalisation of $T_h$, $Z_h$ and $m_h$ in the on-shell scheme, while the fine structure constant $\alpha$ is renormalised in the so-called $G_F$ scheme (cf.~\cite{Denner:1991kt} for a review of the renormalisation of the EW sector of the SM). To~guarantee that the Wilson coefficients or effective couplings obey the usual~SMEFT renormalisation group equations~\cite{Jenkins:2013zja}, the~operator renormalisation is performed in the~$\overline{\rm MS}$ scheme with the renormalisation scale set to $m_h$. Our renormalisation procedure therefore resembles the one employed in the article~\cite{Maltoni:2018ttu} to which we relegate the interested reader for further technical details. 

After renormalisation, we find that the ${\cal O} (\kappa_3 \kappa_4)$ correction associated to the Feynman diagrams of Figure~\ref{fig:diagramsadditional} can be written as 
\beq \label{eq:DeltaF1F2additional}
\Delta {\cal F}_{1,f}^{\kappa_3 \kappa_4} = \frac{\alpha_s}{4 \pi} \, \lambda  \, \kappa_3 \, \frac{\lambda  \, \kappa_4}{(4 \pi)^2} \, g ( \hat s  ) \,. \qquad 
\Delta {\cal F}_{2,f}^{\kappa_3 \kappa_4} = 0 \,,
\eeq
where
\beq \label{eq:gfunction}
g(\hat s) = \frac{18 \hspace{0.25mm} \hat s}{\hat s - m_h^2} \; \tau_t \left [ 1+ (1- \tau_t) \arctan^2 \left ( \frac{1}{\sqrt{\tau_t - 1}} \right ) \right ] \left [  \sqrt{1 - \tau_h} \, \ln \left ( \frac{ \sqrt{1 - \tau_h}  + 1}{\sqrt{1 - \tau_h}  - 1} \right ) + \frac{2 \pi}{\sqrt{3}} - 6 \right ] \,, 
\eeq
and $\tau_a = 4m_a^2/\hat s$ with $a = t, h$ and  it is understood that $\tau_a \to \tau_a - i0$  for analytic continuation. The subscript $f$ in (\ref{eq:DeltaF1F2additional}) indicates that these corrections arise from  two-loop diagrams that factorise into two one-loop graphs. We add that the above result is identical to the one that one obtains if the renormalised $h^3$ vertex is defined via a subtraction at vanishing external momenta as done in~\cite{Liu:2018peg}.

Under the assumption that the operators ${\cal O}_6$ and ${\cal O}_8$ introduced in~(\ref{eq:LSMEFT}) represent the only modifications of the effective one-loop~$h^3$~vertex, we find that the  ${\cal O} (\kappa_4)$ correction  arising from  the graphs displayed in Figure~\ref{fig:diagramsadditional} takes the following form 
\beq \label{eq:DeltaF1F2further}
\Delta {\cal F}_{1,f}^{\kappa_4} = \frac{\alpha_s}{4 \pi} \, \frac{\lambda  \, \kappa_4}{(4 \pi)^2} \, \lambda  \, h ( \hat s  ) \,. \qquad 
\Delta {\cal F}_{2,f}^{\kappa_4} = 0 \,,
\eeq
where
\beq \label{eq:hfunction}
h(\hat s) = -\frac{42 \hspace{0.25mm} \hat s}{\hat s - m_h^2} \; \tau_t \left [ 1+ (1- \tau_t) \arctan^2 \left ( \frac{1}{\sqrt{\tau_t - 1}} \right ) \right ]  \,.
\eeq
We emphasise that the result~(\ref{eq:DeltaF1F2further}) unlike (\ref{eq:DeltaF1F2additional}) is model dependent. This model dependence arises because $\Delta {\cal F}_{1,f}^{\kappa_4}$ receives contributions from Feynman diagrams with quintic scalar self-interactions such as the one shown on the right-hand side of Figure~\ref{fig:diagramsadditional}.  The expressions given in~(\ref{eq:DeltaF1F2further}) and (\ref{eq:hfunction}) correspond to a quintic Higgs self-interaction of the form $V \supset \kappa_5/v \hspace{0.5mm} h^5$ with $\kappa_5 = \lambda \left ( 3 \hspace{0.25mm} \bar c_6 + 14 \hspace{0.25mm} \bar c_8 \right )/4$. Like in~\cite{Maltoni:2018ttu,Borowka:2018pxx} other possible contributions to the coupling modifier $\kappa_5$ such as those that  arise for instance from the higher-dimensional operator ${\cal O}_{10} = -\lambda \hspace{0.5mm} \bar c_{10}/v^6 \, |H|^{10}$ have instead been neglected in our calculation of $\Delta {\cal F}_{1,f}^{\kappa_4}$. This should be contrasted with the analysis~\cite{Liu:2018peg} where it has been assumed that the contributions from higher-dimensional operators other than ${\cal O}_6$ and ${\cal O}_8$ are such that the correction $\Delta {\cal F}_{1,f}^{\kappa_4}$ effectively vanishes. Since (\ref{eq:DeltaF1F2further}) is compared to~(\ref{eq:DeltaF1F2}) parametrically suppressed by a factor of $\lambda/y_t^2 \simeq 0.13$ it turns out that neglecting  the $\bar c_{10}$ contribution in the calculation of $\Delta {\cal F}_{1,f}^{\kappa_4}$ is satisfied in BSM models in which the ratios of Wilson coefficient scale as $\bar c_D/\bar c_{D+2} =  {\cal O} (v^2/\Lambda^2) \ll 1$ or $\bar c_D/\bar c_{D+2}  = {\cal O} (1)$ with $D = 6, 8$. SM extensions that give rise  to $\bar c_{10}/\bar c_{8}  \gg 1$ are, on the other hand, not well described by (\ref{eq:DeltaF1F2further}). To be directly sensitive to the Wilson coefficient $\bar c_{10}$ or equivalently $\kappa_5$, one would need to evaluate quadruple-Higgs production at the tree level or study one-loop effects in triple-Higgs production, and combine the obtained constraints with the measurements of double-Higgs and triple-Higgs production considered in~Section~\ref{sec:numerics}. Such an analysis is, however, beyond the scope of this article.

The two panels in the middle and lower row of Figure~\ref{fig:refimf} show the  real and imaginary parts of~(\ref{eq:gfunction}) and~(\ref{eq:hfunction}), respectively. One sees that the $\hat s$-dependence of both $g(\hat s)$ and $h(\hat s)$ is non-trivial  with  extrema at $\sqrt{\hat s} = 2 m_t \simeq 375 \, {\rm GeV}$ and $\sqrt{\hat s} = 2 \left ( m_h + m_ t \right ) \simeq 600 \, {\rm GeV}$. The observed features will again distort the~kinematic  distributions in $gg \to hh$ production. 

\section{Numerics}
\label{sec:numerics}

In this section we derive limits on the parameters $\Delta \kappa_3$ and $\Delta \kappa_4$ that describe possible modifications of the cubic and quartic Higgs self-couplings with respect to the SM. We consider both double-Higgs and triple-Higgs production at a $27 \, {\rm TeV}$ HE-LHC with an integrated luminosity of $15 \, {\rm ab}^{-1}$ as well as a $100 \, {\rm TeV}$ FCC-pp collider assuming $30 \, {\rm ab}^{-1}$ of   data.  

\subsection{Inclusive double-Higgs and triple-Higgs production}
\label{sec:numerics1}

We begin our study by presenting results for the relevant inclusive production cross sections. In the case of double-Higgs production, we find the following expressions {
\begin{equation} \label{eq:hhxsecs} 
\begin{split}
\sigma \left ( pp \to hh \right )_{\text{HE-LHC}} & = 131 \hspace{1mm} \Big [ 1 - 0.73 \hspace{0.25mm} \Delta \kappa_3  + 1.9 \cdot 10^{-3} \hspace{0.25mm} \Delta \kappa_4 \\[1mm]
& \hspace{1.15cm} + 0.24 \left ( \Delta \kappa_3 \right )^2  + 4.9 \cdot 10^{-4} \hspace{0.25mm} \Delta \kappa_3  \hspace{0.25mm} \Delta \kappa_4 + 2.7 \cdot 10^{-5} \left (  \Delta \kappa_4 \right)^2 \\[1mm] 
& \hspace{1.15cm}  - 1.3  \cdot 10^{-3}  \left ( \Delta \kappa_3 \right )^2 \hspace{0.25mm} \Delta \kappa_4 - 1.8  \cdot 10^{-5}   \hspace{0.25mm} \Delta \kappa_3 \left ( \Delta \kappa_4 \right )^2 \\[1mm]
& \hspace{1.15cm}  + 8.8 \cdot 10^{-6} \left ( \Delta \kappa_3 \right )^2 \left ( \Delta \kappa_4 \right )^2\Big ] \, {\rm fb}  \,, \\[2mm] 
\sigma \left ( pp \to hh \right )_{\text{FCC-pp}} & = 1151 \hspace{1mm} \Big [ 1 - 0.76 \hspace{0.25mm} \Delta \kappa_3  + 2.1 \cdot 10^{-3} \hspace{0.25mm} \Delta \kappa_4 \\[1mm]
& \hspace{1.15cm} + 0.23 \left ( \Delta \kappa_3 \right )^2  + 6.1 \cdot 10^{-4} \hspace{0.25mm} \Delta \kappa_3  \hspace{0.25mm} \Delta \kappa_4 + 3.1 \cdot 10^{-5} \left (  \Delta \kappa_4 \right)^2 \\[1mm] 
& \hspace{1.15cm}  - 1.3  \cdot 10^{-3}  \left ( \Delta \kappa_3 \right )^2 \hspace{0.25mm} \Delta \kappa_4 - 2.1  \cdot 10^{-5}   \hspace{0.25mm} \Delta \kappa_3 \left ( \Delta \kappa_4 \right )^2 \\[1mm]
& \hspace{1.15cm}  + 8.9 \cdot 10^{-6} \left ( \Delta \kappa_3 \right )^2 \left ( \Delta \kappa_4 \right )^2\Big ] \, {\rm fb}  \,.
\end{split}
\end{equation} }
These formulas have been obtained with a customised version of the {\tt POWHEG BOX}~\cite{Alioli:2010xd} implementation of the NLO QCD calculation of double-Higgs production~\cite{Borowka:2016ehy,Borowka:2016ypz,Heinrich:2017kxx} using {\tt PDF4LHC15\_nlo} parton distribution functions~(PDFs)~\cite{Butterworth:2015oua}. Our scale choice is $\mu_R = \mu_F = m_{hh}/2$ with $\mu_R$ and $\mu_F$ denoting the renormalisation and factorisation scale, respectively. As a cross-check we have also derived similar expressions using {\tt MCFM}~\cite{Campbell:2010ff} and {\tt MadGraph5\_aMC@NLO}~\cite{Alwall:2014hca}, finding numerical agreement between all results at~leading order (LO) in QCD.   We note that the SM cross sections that follow from (\ref{eq:hhxsecs}) agree with the central values of the NLO QCD~results quoted in~\cite{Borowka:2016ehy,Borowka:2016ypz,Grazzini:2018bsd} within  a few percent. The observed small differences are in part due  to a slightly different treatment of~$\alpha_s$~in {\tt POWHEG BOX} and the latter~calculations.

In the case of triple-Higgs production, the dependence of the total production cross sections on $\Delta \kappa_3$ and  $\Delta \kappa_4$ instead takes the form 
\begin{align} \label{eq:hhhxsecs} 
\sigma \left ( pp \to hhh \right )_{\text{HE-LHC}} & =   0.44 \hspace{1mm}  \Big  [ 1 - 0.79 \hspace{0.25mm} \Delta \kappa_3  -  0.10  \hspace{0.25mm} \Delta \kappa_4  \nonumber \\[1mm] 
& \hspace{1.25cm} +  0.81 \left ( \Delta \kappa_3 \right)^2  - 0.16 \hspace{0.25mm} \Delta \kappa_3  \hspace{0.25mm} \Delta \kappa_4 + 1.6 \cdot 10^{-2} \left ( \Delta \kappa_4 \right)^2 \nonumber \\[1mm] 
&  \hspace{1.25cm} - \, 0.23 \left ( \Delta \kappa_3 \right )^3  + 4.5 \cdot 10^{-2} \left ( \Delta \kappa_3 \right )^2  \hspace{0.25mm} \Delta \kappa_4 \nonumber \\[1mm] 
&  \hspace{1.25cm}  + 3.5 \cdot 10^{-2} \left ( \Delta \kappa_3 \right )^4    \Big ] \, {\rm fb}  \,, \hspace{6mm} \nonumber \\[2mm] 
\sigma \left ( pp \to hhh \right )_{\text{FCC-pp}} & = 5.1 \hspace{1mm} \Big  [ 1 - 0.67 \hspace{0.25mm} \Delta \kappa_3  -  0.11  \hspace{0.25mm} \Delta \kappa_4 \nonumber \\[1mm] 
& \hspace{1.05cm} +  0.72 \left ( \Delta \kappa_3 \right)^2  - 0.14 \hspace{0.25mm} \Delta \kappa_3  \hspace{0.25mm} \Delta \kappa_4 + 1.6 \cdot 10^{-2} \left ( \Delta \kappa_4 \right)^2 \nonumber \\[1mm] 
& \hspace{1.05cm} - \, 0.20 \left ( \Delta \kappa_3 \right )^3  + 4.0 \cdot 10^{-2} \left ( \Delta \kappa_3 \right )^2  \hspace{0.25mm} \Delta \kappa_4 \nonumber \\[1mm]
& \hspace{1.05cm} + 3.0 \cdot 10^{-2} \left ( \Delta \kappa_3 \right )^4   \Big ] \, {\rm fb}  \,.
\end{align} 
These expressions have been obtained at LO in QCD with the help of {\tt MadGraph5\_aMC@NLO}, taking into account the NLO QCD corrections calculated in~\cite{Maltoni:2014eza} in the form of an overall normalisation.  The used PDF set is again {\tt PDF4LHC15\_nlo}. We add that the $\Delta \kappa_3$ and  $\Delta \kappa_4$ dependence of our FCC-pp result as given in~(\ref{eq:hhhxsecs}) agrees with a similar formula presented in~\cite{Papaefstathiou:2015paa} for the special case~$\Delta \kappa_4 = 6 \Delta \kappa_3$. 


In order to estimate the precision of future hadron colliders in measuring the inclusive double-Higgs production cross section,  we consider  the $b \bar b \gamma \gamma$ final state. This channel has  been identified in the literature~\cite{Contino:2016spe,Goncalves:2018qas,Azatov:2015oxa,Barr:2014sga,He:2015spf,Mangano:2016jyj,Chang:2018uwu} to lead to the best SM signal significance and the highest precision in the measurement of the cubic Higgs self-coupling. At the $27 \, {\rm TeV}$ HE-LHC with $15 \, {\rm ab}^{-1}$ of integrated luminosity the statistical precision of $pp \to hh \to b \bar b \gamma \gamma$ is expected to be around~$14\%$~\cite{Goncalves:2018qas}, while at a $100 \, {\rm TeV}$ FCC-pp collider with $30 \, {\rm ab}^{-1}$  statistical uncertainties in the ballpark of $3\%$ are anticipated~\cite{Contino:2016spe,Goncalves:2018qas,Azatov:2015oxa,Barr:2014sga,He:2015spf,Mangano:2016jyj,Chang:2018uwu}.  Estimating the theoretical uncertainties on the prediction of the signal and the systematic uncertainty on the overall determination of the background rates is more difficult and necessarily has to rely on assumptions. The study of double-Higgs production at approximate next-to-next-to-leading order (NNLO) in QCD~\cite{Grazzini:2018bsd} finds that the inclusive production cross section at $27 \, {\rm TeV}$~($100 \, {\rm TeV}$) is plagued by scale uncertainties of $2.6\%$~($2.1\%$) and uncertainties of $3.4\%$~($4.6\%$) due to unknown  top-quark mass effects. Given these numbers and envisioning that the understanding of top-quark mass effects at NNLO QCD will be drastically improved in the years to come, it seems not  implausible that a total theoretical uncertainty on $\sigma\left  ( pp \to hh \right )$  of order~$3\%$ ($2\%$) may ultimately be achievable at the HE-LHC~(FCC-pp).  A detailed analysis of the systematic uncertainty  on the overall determination of the SM background rates  at a  FCC-pp has been performed in~\cite{Contino:2016spe}. From the results presented in this work one can conclude that the experimental systematic uncertainties may amount to only about~$2\%$, making them subleading compared to other sources of uncertainty. Treating all quoted uncertainties as uncorrelated Gaussian errors then leads to total uncertainties of around $15\%$ and~$5\%$ on the double-Higgs production cross section at the HE-LHC and FCC-pp, respectively. The latter uncertainty estimates will be used in our numerical analysis. 

In the case of triple-Higgs production, $pp \to hhh \to b \bar b b \bar b \gamma \gamma$ is the channel that has obtained the most attention~\cite{Contino:2016spe,Mangano:2016jyj,Papaefstathiou:2015paa,Chen:2015gva,Fuks:2015hna,Kilian:2017nio}. Under optimistic assumptions about the detector performance~(see~\cite{Contino:2016spe,Papaefstathiou:2015paa} for details) these analyses concur that a $100 \, {\rm TeV}$ FCC-pp collider with $30 \, {\rm ab}^{-1}$ of data should be able to exclude triple-Higgs production cross sections that are larger by a factor of~2 than the SM prediction. It may be possible to further improve this 95\%~CL upper limit by considering for instance the $b \bar b b \bar b \tau^+ \tau^-$ final state~\cite{Fuks:2017zkg}, but we will not entertain such a possibility here. A~sensitivity study of triple-Higgs production at the HE-LHC does  to the best of our knowledge not exist. To estimate the sensitivity that a measurement of triple-Higgs production in the $b \bar b b \bar b \gamma \gamma$ channel can achieve at a $27 \, {\rm TeV}$ machine with $15 \, {\rm fb}^{-1}$ of integrated luminosity, we proceed as follows. We generate the dominant background channels, i.e.~$b \bar b b \bar b \gamma \gamma$ and $hhb \bar b$, as well as the triple-Higgs signal at LO in QCD using {\tt MadGraph5\_aMC@NLO}. Our analysis follows the articles~\cite{Contino:2016spe,Papaefstathiou:2015paa} for what concerns $K$-factors, systematic uncertainties, selection cuts   and detector performances such as the $b$-tagging efficiency and the jet-to-photon mis-identification rate. Based on our simulations, we expect $0.2$ and $0.2$ background events from the $b \bar b b \bar b \gamma \gamma$ and $hh b \bar b$ channel, respectively, while for the $pp \to hhh \to  b \bar b b \bar b \gamma \gamma$ signal we predict $0.5$ events within the SM. Using these numbers and calculating the significance from a Poisson ratio of likelihoods modified to incorporate systematic uncertainties on the background, we  find that a $27 \, {\rm TeV}$ HE-LHC with $15 \, {\rm ab}^{-1}$ of data is expected to exclude triple-Higgs production cross sections that are larger  than the SM prediction by a factor of~approximately~11.

\begin{figure}[!t]
\begin{center}
\includegraphics[width=0.95 \textwidth]{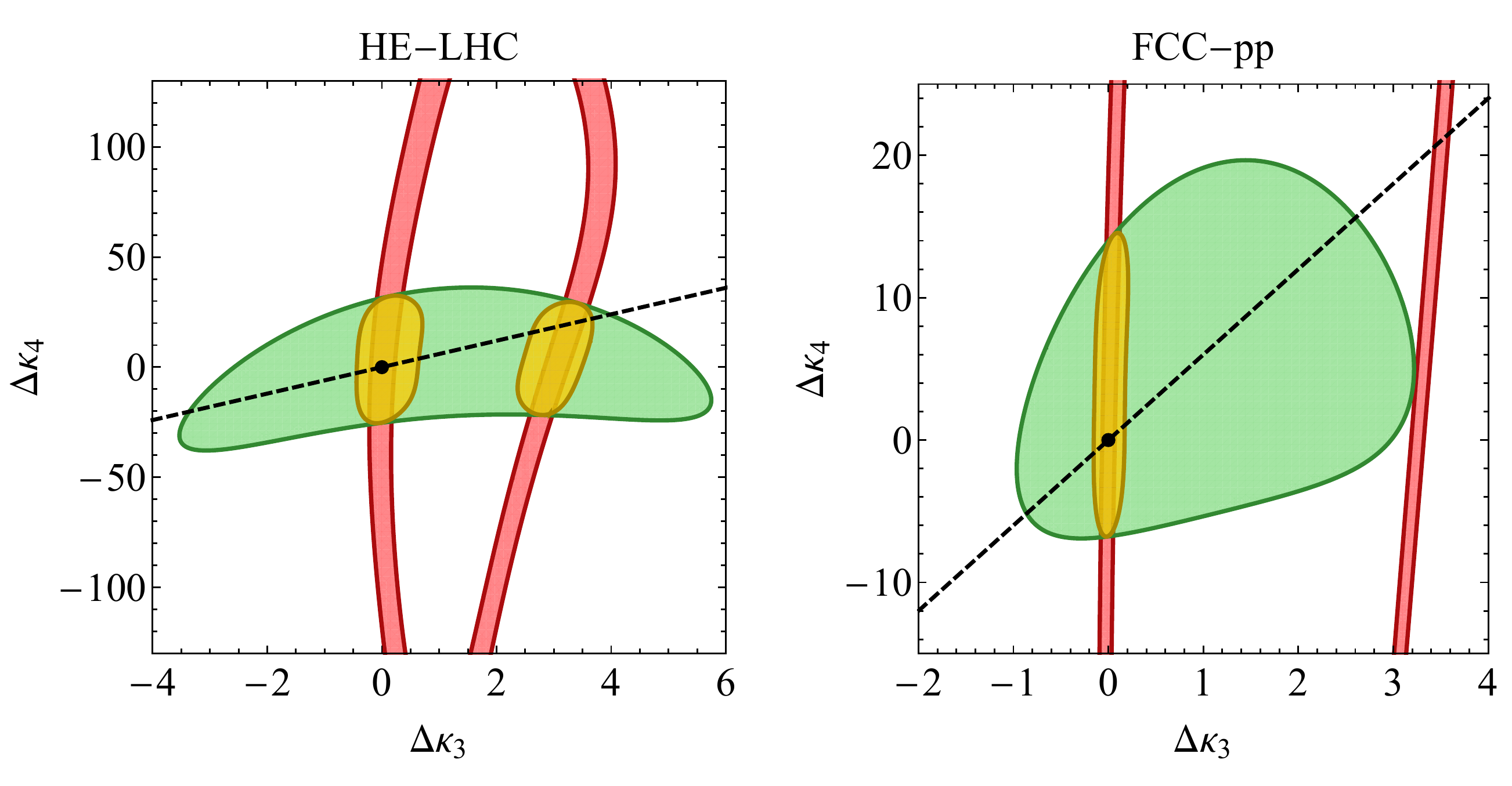} 
\vspace{-2mm}
\caption{\label{fig:xsecprojections} 
Hypothetical constraints in the $\Delta \kappa_3 \hspace{0.25mm}$--$ \hspace{0.25mm} \Delta \kappa_4$ plane.  The red and green contours correspond to the allowed regions in parameter space that arise from double-Higgs and triple-Higgs production, respectively, while the yellow regions are obtained by a combination of the two constraints requiring $\Delta \chi^2 = 5.99$. In~both panels the SM is indicated by the black point and the black dashed line corresponds to $\Delta \kappa_4 = 6 \Delta \kappa_3$. In BSM models that lead to the hierarchy $\bar c_6/\bar c_8  \gg 1$ of Wilson coefficients only $\Delta \kappa_3$ and $\Delta \kappa_4$ values close to the  black dashed  line can be obtained. The results in the left (right) panel have been obtained by assuming that the double-Higgs production cross section has been measured with an accuracy of $15\%$ ($5\%$) at the HE-LHC (FCC-pp). In the case of triple-Higgs production, our projection is instead based on the assumption that cross section values that are larger by a factor of  11~(2) than the SM value are experimentally disfavoured by the HE-LHC~(FCC-pp) at 95\%~CL. See text for further explanations.} 
\end{center}
\end{figure}

The two panels in Figure~\ref{fig:xsecprojections} display the expected exclusion sensitivity in the  $\Delta \kappa_3 \hspace{0.25mm}$--$ \hspace{0.25mm} \Delta \kappa_4$ plane for the $27 \, {\rm TeV}$~HE-LHC~(left) and a $100 \, {\rm TeV}$ FCC-pp collider~(right) with $15 \, {\rm ab}^{-1}$ and $30 \, {\rm ab}^{-1}$ of integrated luminosity, respectively.  The red and green curves illustrate the limits from measurements of the double-Higgs and triple-Higgs production cross sections with the accuracy discussed above, while the yellow regions are the $\Delta \chi^2 = 5.99$ contours (corresponding to a 95\%~CL for a Gaussian distribution) that derive from a combination of the two measurements in the form of a~$\chi^2$~fit. The SM point is indicated by the black dots. One observes that the constraints that arise from the hypothetical measurements of double-Higgs production are bands that form ear-shaped exclusion regions. The widths of the bands are determined by the accuracy of the associated measurement of the inclusive $pp \to hh$ cross section, and as a result the band is narrower by a factor of around~3 for the FCC-pp than the HE-LHC. The shape of the constraints from triple-Higgs production instead depends   on whether a future hardron collider can set an~${\cal O} (10)$ or an~${\cal O} (1)$ bound on the cross section. If, like in the case of the HE-LHC, only rough  limits can be obtained the  triple-Higgs constraint has the form of a  banana  that extends along the $\Delta \kappa_3$ axis, while the allowed region turns out to be oval-shaped,  if a future hadron  collider such as the FCC-pp is able to probe triple-Higgs production cross sections at the SM level. 

Combining the two constraints,  two regions of parameter space remain  viable at the HE-LHC that are centred around $\{0,0\}$ and $\{3,4\}$, respectively. In the case of~$\kappa_3 = 1$, we find  that  the range $\kappa_4 \in [-21, 29]$ is allowed at~95\%~CL. Also notice that the family of solutions $\Delta \kappa_3 = 6 \Delta \kappa_4$~(dashed black line) goes right through the  non-SM region of viable parameters. This implies that measurements of the inclusive double-Higgs and triple-Higgs production cross sections at the HE-LHC are unlikely to be able to tell apart scenarios in which  large modifications of both the $h^3$ and $h^4$ vertices arise from the single operator~${\cal O}_6$ or the two operators ${\cal O}_6$ and  ${\cal O}_8$ $\big($cf. the discussion after~(\ref{eq:kappa3kappa4})$\big)$. The FCC-pp should instead be able to disentangle these two possibilities since it is expected to almost entirely rule out parameters choices in the $\Delta \kappa_3 \hspace{0.25mm}$--$ \hspace{0.25mm} \Delta \kappa_4$ plane that are located close to the point~$\{3,4\}$. Large modifications of the quartic Higgs self-coupling could in such a case only arise from the simultaneous presence of  ${\cal O}_6$ and~${\cal O}_8$. One also sees that the allowed region around the SM-point~$\{0,0\}$ will be largely reduced at the FCC-pp compared to the HE-LHC. Numerically,   the following 95\%~CL range  $ \kappa_4 \in [-5,14]$ is obtained under the assumption that $\kappa_3 = 1$.  The quoted range agrees with the FCC-pp bound on the quartic Higgs self-coupling reported  in~\cite{Contino:2016spe}. 

 Before discussing how modified cubic and quartic Higgs self-couplings impact the kinematic distributions in double-Higgs production, we briefly comment on the  maximal size that the parameters $\Delta \kappa_3$ and $\Delta \kappa_4$ may take.  Under plausible assumptions about the UV structure of the Higgs potential, it has been shown in~\cite{Falkowski:2019tft}  that values $\left | \Delta \kappa_3 \right | \lesssim 4$ are compatible with Higgs and EW precision measurements, direct LHC searches and vacuum stability. The corresponding limit on the modifications of the quartic Higgs self-coupling reads $\left | \Delta \kappa_4 \right | \lesssim 40$, if one assumes hat there is a parametric separation between the EW and new-physics scales, leading to a suppression of operators with dimension higher than six. The quoted upper bounds fall into the same ballpark than the limits obtained in~\cite{Maltoni:2018ttu} from perturbativity considerations (see also \cite{DiLuzio:2017tfn,Chang:2019vez} for related discussions).  Since the mentioned theoretical bounds on the parameters $\Delta \kappa_3$ and $\Delta \kappa_4$ are neither sharp nor model independent, we do not  explicitly indicate them in Figures~\ref{fig:xsecprojections}, \ref{fig:distcprojections} and \ref{fig:allprojections} when reporting the results of our phenomenological $\Delta \chi^2$ fits in the $\Delta \kappa_3 \hspace{0.25mm}$--$ \hspace{0.25mm} \Delta \kappa_4$ plane.

\subsection{Kinematic distributions in double-Higgs production}
\label{sec:numerics2}

In the previous section we have seen that combining the information on the inclusive double-Higgs and triple-Higgs production cross section may not  be able to resolve all ambiguities in the $\Delta \kappa_3$ and $\Delta \kappa_4$ determination --- a feature that is nicely illustrated by  the left panel in Figure~\ref{fig:xsecprojections}.  It is well-known~\cite{Goncalves:2018qas, Azatov:2015oxa,Barr:2014sga,He:2015spf,Contino:2016spe,Mangano:2016jyj,Banerjee:2018yxy,Chang:2018uwu} that precise measurements of differential distributions in double-Higgs production can be used to resolve ambiguities and/or flat directions in the extraction of coupling modifiers or Wilson coefficients, and in the following we will apply this general idea to the case of a simultaneous determination of the cubic and quartic Higgs self-couplings. 

\begin{figure}[!t]
\begin{center}
\includegraphics[clip, trim=2cm 2cm 2cm 2cm,width= \textwidth]{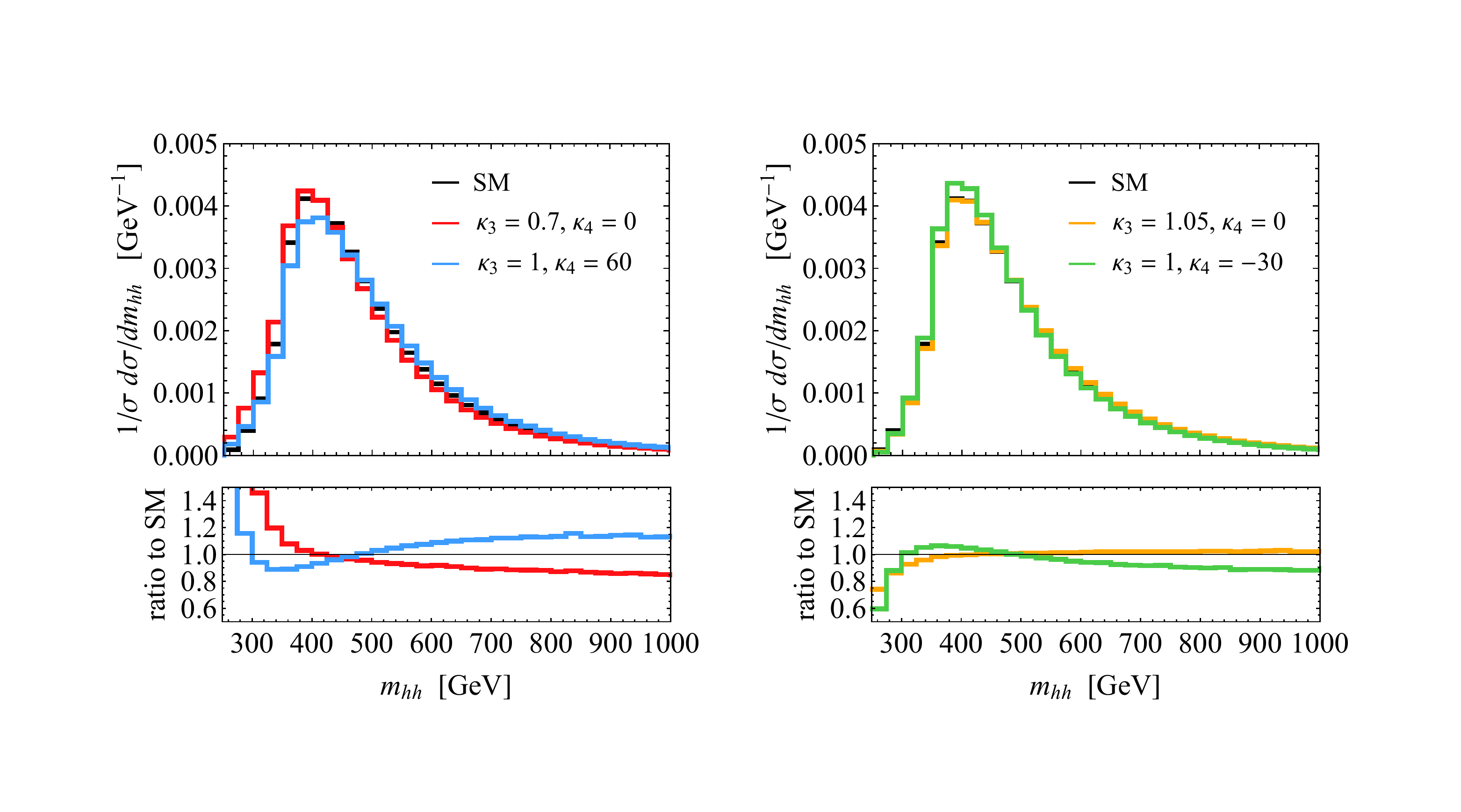} 

\vspace{-2mm}

\includegraphics[clip, trim=2cm 2cm 2cm 2cm,width= \textwidth]{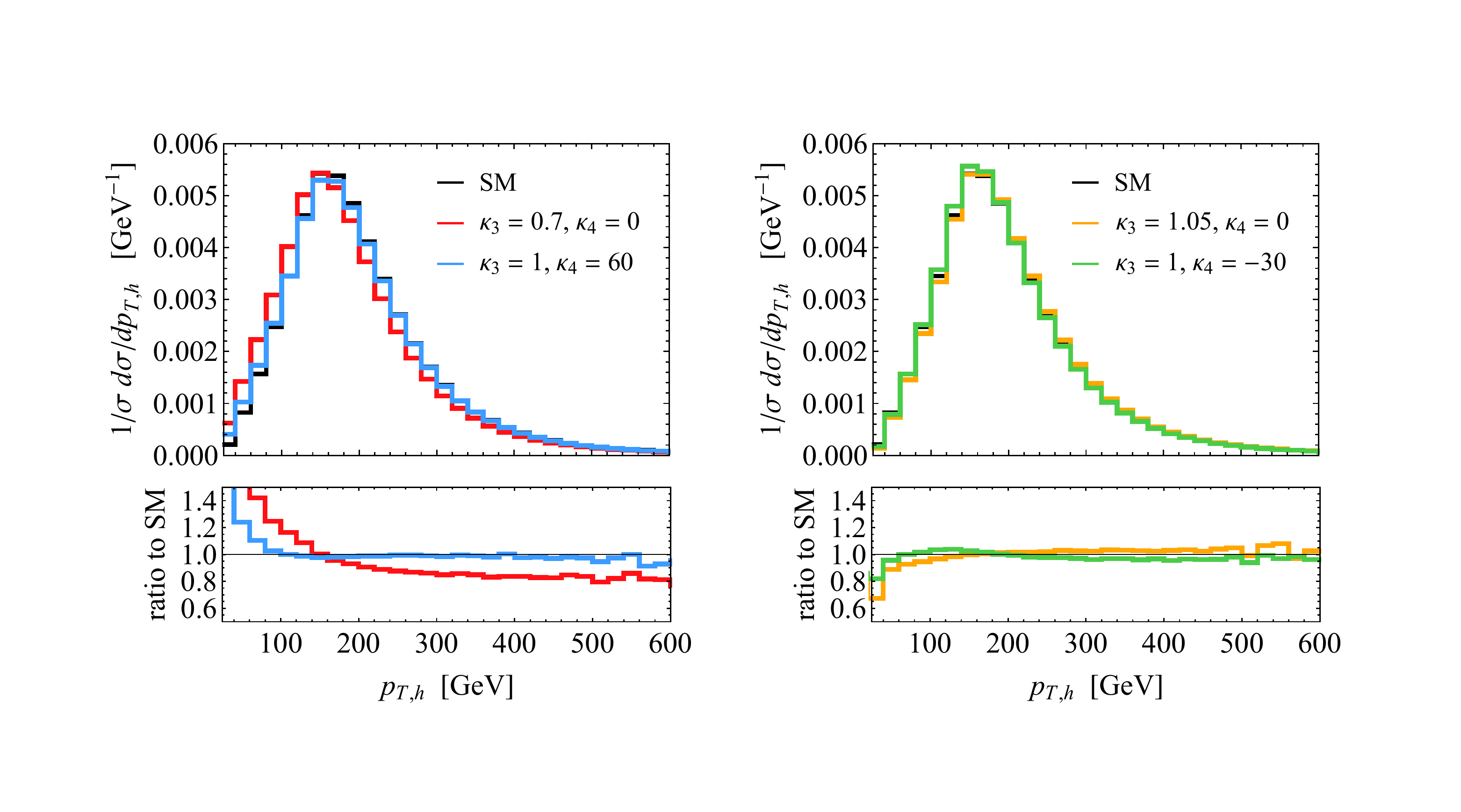} 
\vspace{-4mm}
\caption{\label{fig:distributions} Normalised predictions for the $m_{hh}$ (upper row) and the leading $p_{T,h}$ (lower row) spectrum in $pp \to hh$ production at the $100 \, {\rm TeV}$ FCC-pp.   The black histograms represent the SM distributions, while the coloured curves correspond to different choices of the Higgs boson self-coupling modifiers $\kappa_3$ and~$\kappa_4$. Consult the main text for additional details.} 
\end{center}
\end{figure}

In Figure~\ref{fig:distributions} we depict the differential distributions of two relevant kinematic variables, namely the invariant mass $m_{hh}$ of the di-Higgs system (upper row) and the leading transverse momentum $p_{T,h}$ of the two Higgs bosons (lower row) in $pp \to hh$. The shown results are NLO accurate and, as before, have been obtained with a modified version of {\tt POWHEG BOX} using {\tt PDF4LHC15\_nlo}~PDFs.  They assume $pp$ collisions at a CM energy of $100 \, {\rm TeV}$. The coloured histograms represent  the four choices $\{\kappa_3, \kappa_4 \} = \{0.7,0\},  \{1,60\}, \{1.05,0\},   \{1,-30\}$. The former two parameter combinations lead to enhancements of the inclusive $pp \to hh$ cross section by roughly $30\%$ with respect to the SM, while the latter two choices reduce the double-Higgs production rate by about $-5\%$.  Based on measurements of $\sigma \left (pp \to hh \right )$ the choices $ \{0.7,0\}$ and $\{1,60\}$ $\big( \{1.05,0\}$ and $\{1,-30\} \big)$ are therefore not distinguishable. As can be seen from the four panels in Figure~\ref{fig:distributions}, the  predictions for the normalised $m_{hh}$ and $p_{T,h}$ spectra are however not the same for the two types of $\{\kappa_3, \kappa_4 \}$ sets. Since the distortions in the $p_{T,h}$ distribution turn out to be typically smaller than those in the~$m_{hh}$ spectrum, we will  use the latter kinematic observable in our  shape analysis.

\begin{figure}[!t]
\begin{center}
\includegraphics[width=0.95 \textwidth]{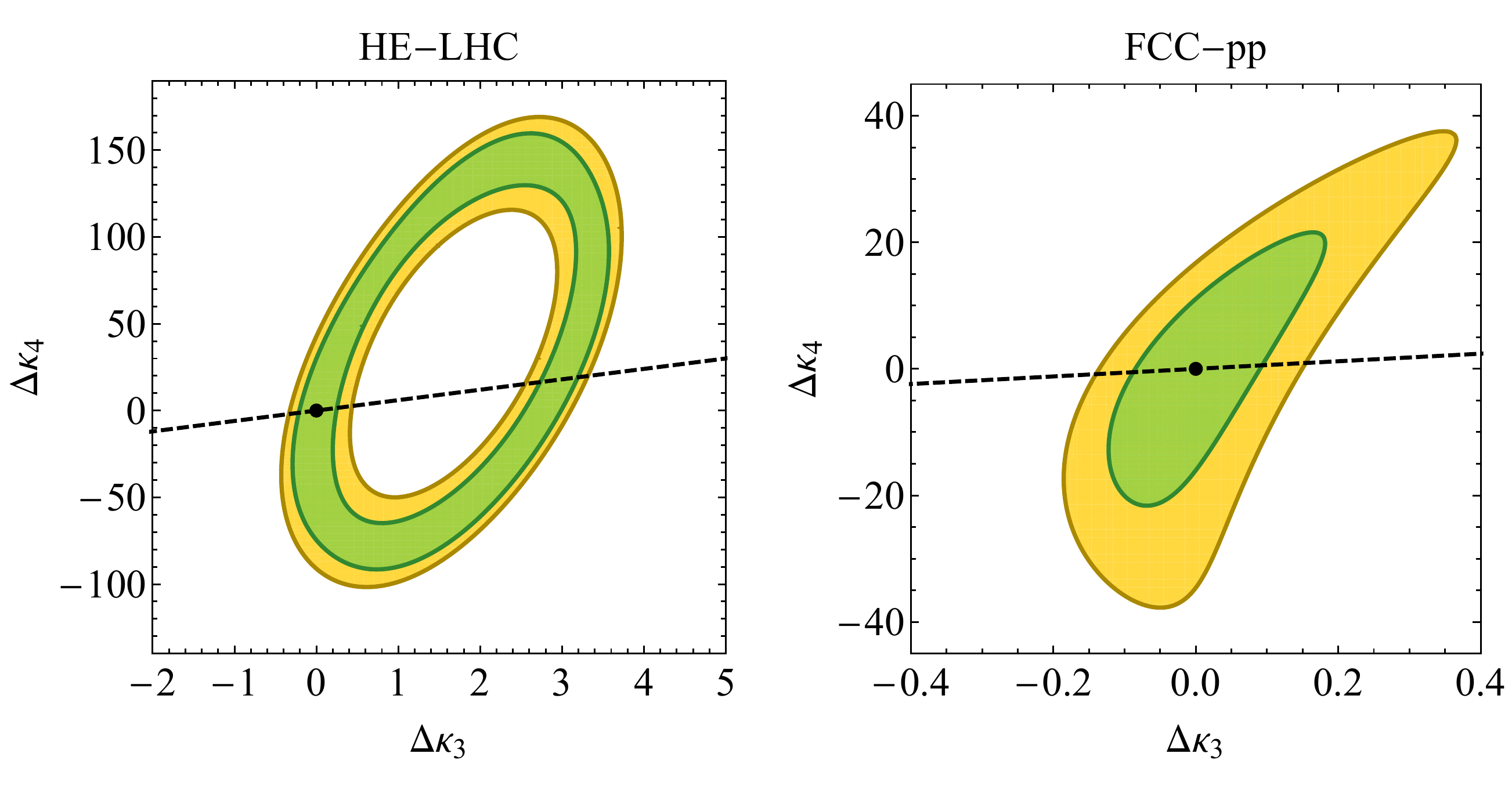} 
\vspace{-2mm}
\caption{\label{fig:distcprojections}
Hypothetical constraints in the $\Delta \kappa_3 \hspace{0.25mm}$--$ \hspace{0.25mm} \Delta \kappa_4$ plane following from a shape analysis of the $m_{hh}$ spectrum in $pp \to hh$ production at the HE-LHC (left panel) and FCC-pp (right panel). The green (yellow) contours correspond to  68\%~CL (95\%~CL) regions. In~both figures the SM is indicated by the black point and the black dashed line represents the family of solutions that satisfy $\Delta \kappa_4 = 6 \Delta \kappa_3$. In SM extensions that give rise to the hierarchy $\bar c_6/\bar c_8 \gg 1$ of dimension-eight and dimension-six contributions only $\Delta \kappa_3$ and $\Delta \kappa_4$ values close to the black dashed line can be accommodated.  For further details consult the text.} 
\end{center}
\end{figure}

The  signal needed to perform the shape analysis is generated at NLO in QCD with {\tt POWHEG~BOX} matched to {\tt  Pythia~8}~\cite{Sjostrand:2007gs,Sjostrand:2014zea} to include parton-shower effects (we use a customised version of the computer code presented in~\cite{Heinrich:2017kxx}). {\tt PDF4LHC15\_nlo}~PDFs are employed and jets ($j$) are reconstructed with {\tt FastJet}~\cite{Cacciari:2011ma}  using an anti-$k_t$ algorithm~\cite{Cacciari:2008gp}. Our analysis then follows~\cite{Goncalves:2018qas}. We demand two $b$-tagged jets ($b$) and two isolated photons ($\gamma$) with the following minimal cuts on the transverse momentum, pseudorapidity and radius separation: $p_{T,x} > 30 \, {\rm GeV}$, $|\eta_x| > 2.5$ and $\Delta R_{xy} > 0.4$ for $x,y = j,b,\gamma$. A flat $b$-tagging efficiency of~$70\%$, and mis-tag rates of $15\%$ for charm quarks and $0.3\%$ for light flavours are adopted. Events with more than three jets are vetoed, and the requirements $|m_{b \bar b} - m_h| < 25 \, {\rm GeV}$,  $|m_{\gamma \gamma} - m_h| < 1 \, {\rm GeV}$ and $m_{hh} > 400 \, {\rm GeV}$ are imposed as a final selection. The obtained $m_{hh}$ distributions have then been binned into bins of $25 \, {\rm GeV}$. Our shape fit includes the statistical uncertainties in each bin as well as theoretical and experimental systematic uncertainties of $3\%$ ($2\%$) and $2\%$ ($2\%$) at HE-LHC (FCC-pp), respectively. The quoted  uncertainties have been treated as uncorrelated Gaussian errors  in the $\chi^2$ fit. We emphasise that  our  fit does not consider the impact of backgrounds, but we have verified that with the described methodology we are able to  reproduce the CL-level curves presented in~\cite{Goncalves:2018qas} for both the HE-LHC and FCC-pp quite well. This agreement gives us confidence that our simplified approach is able to mimic  the more sophisticated analysis~\cite{Goncalves:2018qas} that includes a simulation of all relevant SM backgrounds.

The results of our $m_{hh}$ shape analysis are shown in Figure~\ref{fig:distcprojections}. The green (yellow) regions are  the $\Delta \chi^2 = 2.28$ ($\Delta \chi^2 = 5.99$) contours, corresponding to 68\%~CL (95\%~CL)  limits for a Gaussian distribution. In~both panels the SM point is indicated by a black dot and the black dashed line illustrates the equality $\Delta \kappa_4 = 6 \Delta \kappa_3$.  From the panel on the left-hand side one sees that at the HE-LHC a shape analysis of the $m_{hh}$ distribution in $pp \to hh$ will not allow one to exclude choices in the $\Delta \kappa_3 \hspace{0.25mm}$--$ \hspace{0.25mm} \Delta \kappa_4$  plane around $\{3, 4\}$,~i.e.~parameters that survive a combination of the measurements of the inclusive  double-Higgs and triple-Higgs production cross sections  (see the left panel in Figure~\ref{fig:xsecprojections}). For~$\kappa_3 =1$ we find  the following 95\%~CL range  $\kappa_4 \in [-82, 37]$.  As shown in the right panel in Figure~\ref{fig:distcprojections}, at the FCC-pp the constraints in the $\Delta \kappa_3 \hspace{0.25mm}$--$ \hspace{0.25mm} \Delta \kappa_4$  plane that follow from a $m_{hh}$ shape analysis are expected to improve noticeably compared to the corresponding HE-LHC limits. Differential measurements of $pp \to hh$ at the FCC-pp alone will in consequence be able to distinguish scenarios in which  large modifications of both the $h^3$ and $h^4$ interactions arise from the operator ${\cal O}_6$ or a combination of ${\cal O}_6$ and  ${\cal O}_8$ $\big($cf.~the text after (\ref{eq:kappa3kappa4})$\big)$. Assuming again that $\kappa_3 =1$, the 95\%~CL range for the parameter $\kappa_4$ reads $\kappa_4 \in [-22, 15]$.  Profiling over $\kappa_3$ by means of the profile likelihood ratio~\cite{Cowan:2010js}, we obtain the following 95\%~CL bound  $\kappa_4 \in [-89, 159]$ and $\kappa_4 \in [-19, 21]$ at the HE-LHC and the FCC-pp, respectively. Our FCC-pp constraints on $\kappa_4$ are comparable to those that have been derived in the analysis~\cite{Borowka:2018pxx}.

\subsection{Global fit at the HE-LHC and a FCC-pp}
\label{sec:numerics3}

\begin{figure}[!t]
\begin{center}
\includegraphics[width=0.95 \textwidth]{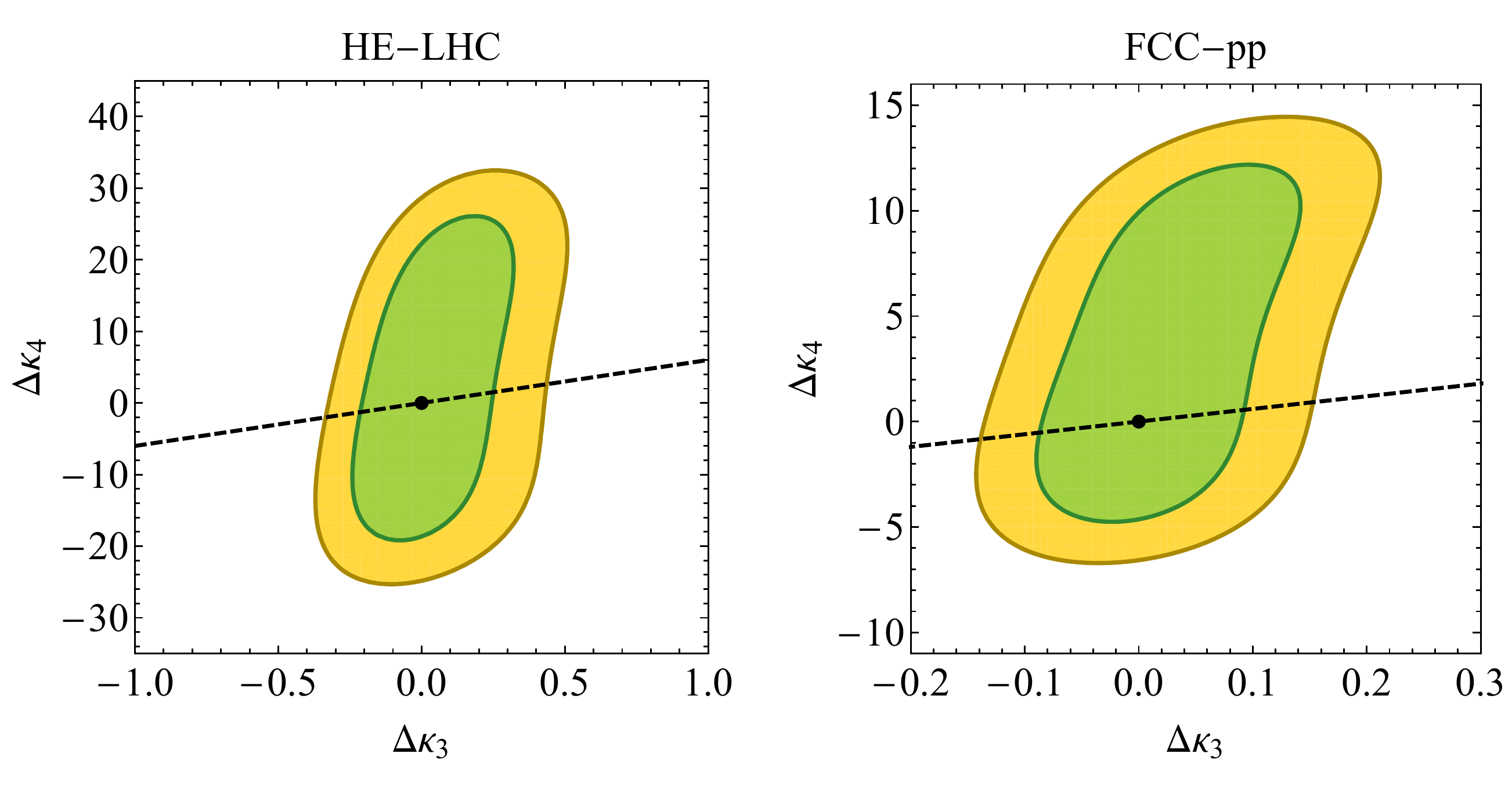} 
\vspace{-2mm}
\caption{\label{fig:allprojections} 
Hypothetical constraints in the $\Delta \kappa_3 \hspace{0.25mm}$--$ \hspace{0.25mm} \Delta \kappa_4$ plane following from a combination of a shape analysis of the $m_{hh}$ spectrum in $pp \to hh$ production and a measurement of the inclusive production cross section of $pp \to hhh$. The green (yellow) contours correspond to  68\%~CL (95\%~CL) regions and the left (right) panel shows the HE-LHC (FCC-pp) projections. The SM solution is indicated by the black point and the black dashed line represents the parameter choices satisfying $\Delta \kappa_4 = 6 \Delta \kappa_3$.   Only $\Delta \kappa_3$ and $\Delta \kappa_4$ values with $\Delta \kappa_4 \simeq 6 \Delta \kappa_3$ can be obtained in BSM models that give rise to $\bar c_6/\bar c_8  \gg 1$.  See text for additional details.} 
\end{center}
\end{figure}

The full potential of the HE-LHC and the FCC-pp in constraining simultaneously the coupling modifications $\kappa_3$ and $\kappa_4$ can be assessed by combining the information on the differential measurements of $pp \to hh$ with the expected accuracies in the determination of the inclusive  $pp \to hhh$ production cross section. The outcome of such an exercise is presented in Figure~\ref{fig:allprojections}. Here the green~(yellow) contours correspond to  68\%~CL (95\%~CL) regions, while the black dots represent the SM point  and the black dashed lines illustrate parameter choices of the form $\Delta \kappa_4 = 6 \Delta \kappa_3$.   {Numerically, we find that for $\kappa_3 = 1$, the 95\%~CL bounds on $\kappa_4$ from a global analysis of differential double-Higgs and inclusive triple-Higgs data at the HE-LHC (FCC-pp) is $ \kappa_4 \in [-21, 27]$ ($ \kappa_4 \in [-5, 12]$). Notice that these limits represent an  improvement of the bounds derived in Section~\ref{sec:numerics1} based on inclusive measurements alone.  Profiling instead over $\kappa_3$, the following  95\%~CL bounds are obtained $\kappa_4 \in [-17, 26]$ and $\kappa_4 \in [-3, 13]$. 

\subsection[Comparison between sensitivities of future $pp$ and $e^+ e^-$ machines]{Comparison between sensitivities of future $\bm{pp}$ and $\bm{e^+ e^-}$ machines}
\label{sec:numerics4}

The constraints on the cubic and quartic Higgs self-couplings that high-energy $e^+ e^-$ machines may be able to set have been studied recently in~\cite{Maltoni:2018ttu,Liu:2018peg}. Both articles have performed global fits assuming various  ILC and CLIC setups to determine the allowed modifications of the cubic and quartic Higgs self-couplings. Under the assumption of $\kappa_3 = 1$, the article~\cite{Maltoni:2018ttu}  finds for instance for an ILC with a CM energy of $500 \, {\rm GeV}$, polarisations of $P(e^-,e^+) = (-0.8, 0.3)$ and an integrated luminosity of $4 \, {\rm ab}^{-1}$ (ILC-500) the following 95\%~CL bound $\kappa_4 \in [-11, 13]$.  At CLIC with a CM energy of $3000 \, {\rm GeV}$, polarisations of $P(e^-,e^+) = (-0.8, 0.0)$ and an integrated luminosity of $2 \, {\rm ab}^{-1}$~(CLIC-3000) the corresponding limit is said to be  $\kappa_4 \in [-5, 7]$. The constraints presented in~\cite{Liu:2018peg} are less stringent than those obtained in~\cite{Maltoni:2018ttu}.   Taking the limits given in~\cite{Maltoni:2018ttu} at face value and comparing them to the bounds presented in the last subsection, suggests that the HE-LHC has a weaker sensitivity to modifications of the quartic Higgs self-coupling than the ILC-500. On the other hand, the FCC-pp reach seems to be better than that of the  ILC-500, and roughly comparable to the CLIC-3000 potential.  

\section{Conclusions}
\label{sec:conclusions}

In this work, we have investigated the possibility to  constrain the quartic Higgs  self-interactions indirectly through precise measurements of double-Higgs production at future hadron colliders. We have first presented the results of a calculation of the two-loop contributions to the $gg \to hh$ amplitudes that involve a modified $h^4$ vertex. Our  results have been obtained in numerical form with the help of {\tt pySecDec}~\cite{Borowka:2012yc,Borowka:2015mxa,Borowka:2017idc} and have been implemented into  {\tt POWHEG BOX}~\cite{Alioli:2010xd}. Combining  the  two-loop EW corrections calculated here with the~${\cal O} (\alpha_s^2)$~matrix elements computed  in~\cite{Borowka:2016ehy,Borowka:2016ypz,Heinrich:2017kxx}, we are able to  predict the cross section and the most important distributions for double-Higgs production at NLO in QCD,  including arbitrary modifications of both the $h^3$ and the~$h^4$ coupling.  

Based on our results, we have then performed an exploratory study of the sensitivity of the $27 \, {\rm TeV}$~HE-LHC and a $100 \, {\rm  TeV}$ FCC-pp in constraining simultaneously the cubic and quartic Higgs self-couplings by measurements of double-Higgs and triple-Higgs production in gluon-fusion.  In a first step, we have considered only measurements of total rates. In the case of the HE-LHC with $15 \, {\rm ab}^{-1}$ of integrated luminosity, we have found that a combined fit to $\sigma \left ( pp \to hh \right )$ and $\sigma \left ( pp \to hhh \right )$ will have a two-fold ambiguity  in the $ \kappa_3 \hspace{0.25mm}$--$ \hspace{0.25mm} \kappa_4$ plane with a family of solutions located either  around $\{1,1\}$ or in the vicinity of~$\{4,5\}$.} The resulting  bounds on possible modifications of the quartic Higgs self-coupling turn out to be generically weak. For instance,  for~$\kappa_3 = 1$ we found that $\kappa_4$ values in the range $\kappa_4 \in [-21, 29]$ are allowed at~95\%~CL.  Due to its significantly improved sensitivity to triple-Higgs production, a FCC-pp with $30 \, {\rm ab}^{-1}$ of data should be able to resolve the aforementioned degeneracy by reducing the viable parameter space to a stripe in the $ \kappa_3 \hspace{0.25mm}$--$ \hspace{0.25mm} \kappa_4$ plane with SM-like cubic Higgs self-couplings. Numerically, we found that for $\kappa_3 = 1$, the range of $ \kappa_4 \in [-5,14]$ is allowed at 95\%~CL.  Our limit agrees with the FCC-pp bound on the $h^4$ coupling quoted in~\cite{Contino:2016spe}. 

Given that precise measurements of differential distributions in double-Higgs production are known~\cite{Goncalves:2018qas, Azatov:2015oxa,Barr:2014sga,He:2015spf,Contino:2016spe,Mangano:2016jyj,Banerjee:2018yxy,Chang:2018uwu} to be able to resolve degeneracies in the extraction of coupling modifiers or Wilson coefficients, we have in a second step performed a shape analysis to determine the allowed regions in the $ \kappa_3 \hspace{0.25mm}$--$ \hspace{0.25mm} \kappa_4$ plane. We have considered both the $m_{hh}$ spectrum and the leading $p_{T,h}$ distribution and found the former observable to have more discriminating power in a simultaneous extraction of  the cubic and quartic Higgs self-couplings.  From our $m_{hh}$ shape analysis it follows   that at the HE-LHC with $15 \, {\rm ab}^{-1}$ of data it should be possible to constrain $\kappa_4$ to the 95\%~CL range  $\kappa_4 \in [-82, 37]$, if one assumes that $\kappa_3 = 1$. The corresponding constraint at a FCC-pp with $30 \, {\rm ab}^{-1}$ of integrated luminosity turns out to be $\kappa_4 \in [-22, 15]$. The 95\%~CL bounds $\kappa_4 \in [-89, 159]$ and $\kappa_4 \in [-19, 21]$ instead apply if one profiles over $\kappa_3$. The obtained limits show that differential measurements in the $pp  \to hh$ channel alone are expected to lead compared to measurements of the inclusive $pp \to hhh$ cross sections to notable weaker  determinations of the quartic Higgs self-coupling at both the HE-LHC and a~FCC-pp. The same conclusion has been drawn in~\cite{Borowka:2018pxx} for what concerns the FCC-pp.

 To assess the full potential of the HE-LHC and the FCC-pp in constraining simultaneously the coupling modifiers $\kappa_3$ and $\kappa_4$, we have combined the differential measurements of $pp \to hh$ with the inclusive measurements of $pp \to hhh$.  Our global analysis demonstrates that under the assumption $\kappa_3 = 1$, one can expect to obtain a 95\%~CL bound on $\kappa_4$ at the HE-LHC (FCC-pp) of $ \kappa_4 \in [-21, 27]$ ($ \kappa_4 \in [-5, 12]$). By profiling over $\kappa_3$, we  arrived at $\kappa_4 \in [-17, 26]$ and $\kappa_4 \in [-3, 13]$. The former bounds can be compared to the hypothetical constraints of the Higgs self-couplings that high-energy $e^+ e^-$ machines might be able to set~\cite{Maltoni:2018ttu,Liu:2018peg}. For example, the ILC-500 (CLIC-3000) is expected to be able to set  the 95\%~CL bound $ \kappa_4 \in [-11,13]$ ($ \kappa_4 \in [-5,7]$)~\cite{Maltoni:2018ttu},  assuming that $\kappa_3 =1$. These numbers indicate that the HE-LHC should have a weaker sensitivity to modified quartic Higgs self-interactions than ILC-500. A FCC-pp and CLIC-3000 can, however, be expected to have roughly similar potentials in constraining the coupling modifier $\kappa_4$.

\acknowledgments 
We thank  Fady Bishara for useful discussions  about profile likelihood ratios, Sophia Borowka for her patient help with {\tt pySecDec} and are grateful to Paolo~Torrielli and Marco~Zaro for their assistance with {\tt MadGraph5\_aMC@NLO}. We express gratitude to Fady Bishara and Marek Sch\"onherr for trying to be good {\tt Sherpa}s~\cite{Gleisberg:2008ta} (so that we could lean back and enjoy).  We~finally would like to express our gratitude to Kunfeng Lyu for drawing our attention to the ${\cal O} (\kappa_3 \kappa_4)$ contributions~(\ref{eq:DeltaF1F2additional}) that have not been considered in the first version of this article.  WB~has been supported by the ERC Consolidator Grant HICCUP (614577), UH acknowledges the   hospitality  and support of the CERN Theoretical Physics Department and LR has been supported by the ERC Starting Grants PDF4BSM~(335260) and REINVENT (714788).


%

\end{document}